\documentclass{iopjournal}

\usepackage{dcolumn}
\usepackage{bm}

\pdfoutput=1
\usepackage[utf8]{inputenc}
\usepackage[english]{babel}
\usepackage[T1]{fontenc}
\usepackage{amsmath}
\hypersetup{colorlinks,allcolors=black}
\usepackage{braket}
\usepackage{tikz}
\usetikzlibrary{positioning}
\usetikzlibrary{shapes.geometric}
\usepackage{lipsum}
\usepackage{caption}
\usepackage{subcaption}
\usepackage[export]{adjustbox}
\usepackage{algorithm}
\usepackage{algpseudocode}

\usepackage{color,soul}
\usepackage{amssymb}
\usepackage[version=4]{mhchem}

\begin{document}

\articletype{Paper}

\title{Optimised Fermion-Qubit Encodings for Quantum Simulation with Reduced Circuit Depth}


\author{Michael Williams de la Bastida$^{1, 2,*}$\orcid{0000-0000-0000-0000}, Thomas M. Bickley$^{1,2}$\orcid{0000-0000-0000-0000} and Peter V. Coveney$^{1,2}$\orcid{0000-0000-0000-0000}}

\affil{$^1$Centre for Computational Science, Department of Chemistry, University College London, London WC1H 0AJ, United Kingdom}

\affil{$^2$UCL Centre for Advanced Research Computing, Gower Street, London WC1E 6BT, United Kingdom}

\affil{$^*$Author to whom any correspondence should be addressed.}

\email{michael.williams.20@ucl.ac.uk}


\begin{abstract}
Simulation of fermionic Hamiltonians with gate-based quantum computers requires the selection of an encoding from fermionic operators to quantum gates, the most widely used being the Jordan-Wigner transform. Many alternative encodings exist, with quantum circuits and simulation results being sensitive to choice of encoding, device connectivity and Hamiltonian characteristics. Non-stochastic optimisation of the ternary tree class of encodings to date has targeted either the device or Hamiltonian. We develop a deterministic method which optimises ternary tree encodings without changing the underlying tree structure. This enables reduction in Pauli-weight without ancillae or additional swap-gate overhead. We demonstrate this method for a variety of encodings, including those which are derived from the qubit connectivity graph of a quantum computer. Numerical results for a suite of standard encoding methods applied to water in the \emph{STO-3G} basis indicate that our method reduces qDRIFT circuit depths on average by $24.7\%$ and $26.5\%$ for untranspiled and transpiled circuits respectively.
\end{abstract}

\section{Introductory Material}
\label{sec:intro}

Chemistry, materials and life sciences are among the primary uses of high-performance computing (HPC) resources. The utility of chemistry simulation to a wide range of industries has only increased as more powerful computing resources have become available, enabling more varied and larger systems to be studied. Scaling of classical HPC resources will not continue to provide access to more accurate simulation of larger systems indefinitely, however. For example, when seeking to solve the electronic structure Hamiltonian, algorithmic complexity presents a block to progress, in the worst case for an exact solution, scaling factorially \cite{eriksen_shape_2021}.

Quantum computing offers a route by which such simulations could be carried out for much larger systems \cite{leimkuhler_exponential_2025}. In an early call for research into quantum information processing, many-body physics was the defining use case envisaged for quantum computers \cite{feynman_simulating_1982}. Before the advantages of quantum computing can be realised however, numerous technical challenges must be addressed. Many of these relate to the limitations of currently available quantum computing hardware. Low qubit counts restrict the systems which may be represented to only the smallest. Errors in gate implementation and readout limit the number of operations. Low coherence times restrict the duration of quantum algorithms. While remedies to these are sought through improvements to quantum processors, algorithmic methods can be used ensure that the maximum possible benefit is obtained for a given problem and device. In turn, research and development of devices can focus on enabling the most effective algorithms. This \textit{co-design} approach is expected to be vital to the realisation of quantum advantage \cite{BuzzStrategicPaths2024}.

In recent years there have been many publications on the generation and optimisation of fermion-qubit encodings, which are required to simulate fermionic systems with qubits, with much of this focused on the ternary tree (TT) class of encodings \cite{li_huffman-code-based_2025, liu_hatt_2025, miller_bonsai_2023, miller_treespilation_2024}. Two distinct aims can be seen in the literature. First, by starting with a specific fermionic Hamiltonian, to construct an encoding that returns a qubit Hamiltonian which is in some sense optimal \cite{li_huffman-code-based_2025, liu_hatt_2025, chien_optimizing_2022, miller_treespilation_2024, yu_clifford_2025}. Secondly, by starting with the connectivity graph of qubits in a quantum processing unit (QPU), to construct a TT subgraph of this \cite{miller_bonsai_2023, miller_treespilation_2024}. Methods of the first kind may return encodings with an underlying tree structure that is not compatible with the QPU to be used, requiring additional expensive swap gates to enable interaction between qubits which neighbour each other in the encoding tree, but not on the device. Methods of the second kind do not exploit the structure of the Hamiltonian to be simulated, requiring additional gates to implement operators. Work to date which has sought to combine these aims has used stochastic optimisation methods such as simulated annealing \cite{yu_clifford_2025, miller_treespilation_2024}. These methods are not guaranteed to find global minima, and are sensitive to the initial parameter setting.

The purpose of the present paper is to develop and demonstrate a deterministic optimisation method which leaves the underlying graph structure of an encoding intact, allowing for the construction of encodings which are optimised for both device and Hamiltonian. We demonstrate that this method has broad applicability, low computational cost and favourable scaling. Further, we show that substantial reductions in circuit depth for the stochastic quantum simulation algorithm `qDRIFT' are obtained for all encodings.

We begin Section \ref{sec:encoding} by introducing the most commonly used class of encodings, Majorana-string encodings, before describing their generation from ternary trees. Section \ref{sec:optimisation} presents cost-functions and the degrees of freedom in these encodings over which optimisation will be performed. Our method, Topology-Preserving Hamiltonian Adaptive Ternary Tree (TOPP-HATT) is described in Section \ref{sec:TOPP-HATT}. Results are presented in Section \ref{sec:results}, including demonstrations of our method with a variety of standard encodings, and its application to reduce the depth of qDRIFT circuits, which implement the Hamiltonian time evolution operator.

\section{Fermion to Qubit Encodings}
\label{sec:encoding}


Given a fermionic Hamiltonian in second quantised form, each term is composed of some combination of fermionic creation and annihilation operators, $a_{i}^{\dagger}$ and $a_{i}$ respectively. To implement simulations of fermionic systems, a fermion-qubit encoding which maps fermionic operators to a set of qubit-operators is required. 

In general an encoding is an isometry from the anti-symmetric Fock space of N fermions in M modes $\mathcal{F^{-}_{N,M}}$ into a complex Hilbert space of $m$ qubits $\mathcal{H}^{q}_{m} \equiv (\mathbb{C}^{2})^{\otimes m}$ \cite{bravyi_tapering_2017}. The eigenvalues in any suitable encoding must be the same as those of the fermionic system, but the eigenvectors need not be \cite{ReducingEntanglementPhysically2024}. As such, there are many possible encoding schemes. 
An additional practical concern constrains the useful encodings, in that we require an encoding for which it is straightforward to prepare the vacuum state $\ket{}_{f}$. Encodings which map the computational basis state $\ket{0}^{\otimes m}$ to the vacuum state $\ket{}_{f}$ are called \emph{vacuum preserving}.
The focus of this work is vacuum-preserving Majorana-string encodings.

\subsection{Majorana-string Encodings}
\label{sec:mse}

Fermionic creation and annihilation operators, $a^{\dagger}$ and $a$, may be decomposed in terms of Majorana operators:
\begin{equation}
\gamma_{\beta_{2j}}= a_{\alpha_{j}}+a_{\alpha_{j}}^{\dagger},
\end{equation}

\begin{equation}
    \gamma_{\beta_{2j+1}}= -i \cdot (a_{\alpha_{j}}-a_{\alpha_{j}}^{\dagger}).
\end{equation}

Inverting these, we have:
\begin{equation}
    a_{j}=\frac{1}{2}(\gamma_{2j}+i\gamma_{2j+1}),
    \label{eq:annihiliation}
\end{equation}
\begin{equation}
    a^{\dagger}_{j} = \frac{1}{2}(\gamma_{2j}-i\gamma_{2j+1}).
    \label{eq:creation}
\end{equation}

For an anti-symmetric Fock space of $M$ modes and $N \le M$ electrons, $\mathcal{F}_{N \le M,M}^{-}$, this requires a set of $2M$ Majorana operators. These Majorana operators can then be mapped to a set of Pauli strings given the following conditions: \cite{miller_bonsai_2023}
\begin{enumerate}
    \item Each Majorana operator is mapped to a Pauli string,
        $$m_{j} \to S_{j}\in \{S\}\ \forall j\in\{0,\dots,2M-1\}.$$
    \item The Pauli strings satisfy the anti-commutation relation: 
    $$\{  S_{i},S_{j} \}=2\delta_{ij} \mathbb{I}.$$
    \item The Pauli strings are linearly independent.
    \item The Pauli strings are algebraically independent: \newline For all unequal subsets $A \subseteq S$ and $B \subseteq S$ such that $A \ne B$, $\prod_{S_{i}\in A}S_{i}\propto \prod_{S_{j}\in B}S_{j}$ is not fulfilled. 
\end{enumerate}

Pauli strings are composed of the operators $\{\hat{X},\hat{Y},\hat{Z},\hat{I} \}$, so they obey condition $2$ if each string has an odd number of non-trivial overlaps (NTO) with each of the others. An overlap, where two Pauli strings act on the same qubit, is non-trivial if the operators do not commute.

Listing each of the $2M$ operators as Pauli strings of length M (using $\hat{I}$ to pad as needed), the Jordan-Wigner encoding for four modes becomes:

\begin{equation}
    \begin{matrix}
    \mathcal{E}_{JW} =& 
        \{XIII, YIII, ZXII, ZYII,\\& ZZXI, ZZYI, ZZZX, ZZZY\}.
    \end{matrix}
    \label{eq:jw4}
\end{equation}

\begin{figure*}[h!]
    \centering
    \begin{subfigure}[b]{0.3\textwidth}
        \includegraphics[width=\textwidth]{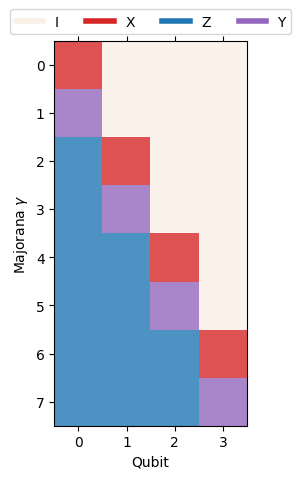}
        \caption{4-Mode Jordan-Wigner}
        \label{fig:jw-ops}
    \end{subfigure}
    \hspace{0.1\textwidth}
    \begin{subfigure}[b]{0.3\textwidth}
        \includegraphics[width=\textwidth]{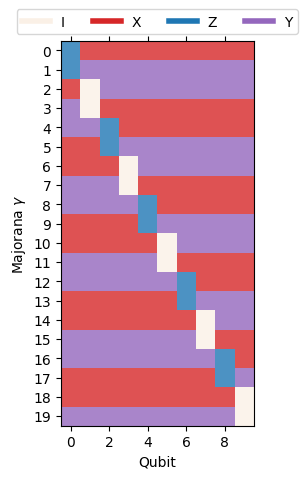}
        \caption{MaxNTO Encoding}
        \label{fig:maxnto-ops}
    \end{subfigure}
    \caption{Graphical representation of Majorana-string encodings.}{a) JW encoding for four modes. b) 10 mode, 9-NTO Max-NTO encoding. Each row represents a Pauli-string mapped to a Majorana operator, $\hat{X}$ operators are shown as red squares in their respective positions, while $\hat{Y}$ and $\hat{Z}$ are purple and blue respectively.
    }
    \label{fig:ms-ops}
\end{figure*}
The Jordan-Wigner encoding has exactly 1-NTO between each Pauli string, which can be easily verified by examining the operators directly. As encodings become larger, and their structures more complex, their representations as Pauli-strings become difficult to read, and the relationships between operators are not always obvious. Going forward, we represent Majorana-string encodings with colourised matrices; for example the Jordan-Wigner encoding of Equation \ref{eq:jw4} is shown in Figure \ref{fig:jw-ops}. The utility of this visualisation becomes more obvious with larger systems, or when considering the permutations of symmetries present in an encoding, as we shall do in Section \ref{sec:symmetry}.
\begin{figure*}[h!]
    \centering
    \begin{subfigure}[b]{0.3\textwidth}
        \includegraphics[width=\textwidth]{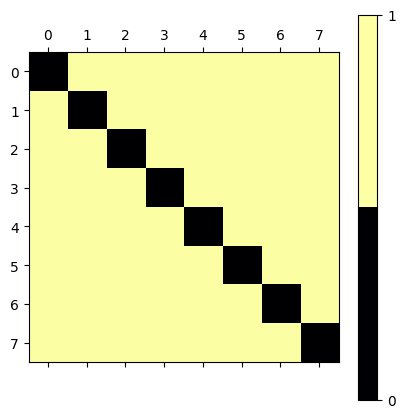}
        \caption{4-Mode Ternary Tree}
        \label{fig:jw-ntos}
    \end{subfigure}
    \hspace{0.1\textwidth}
    \begin{subfigure}[b]{0.3\textwidth}
        \includegraphics[width=\textwidth]{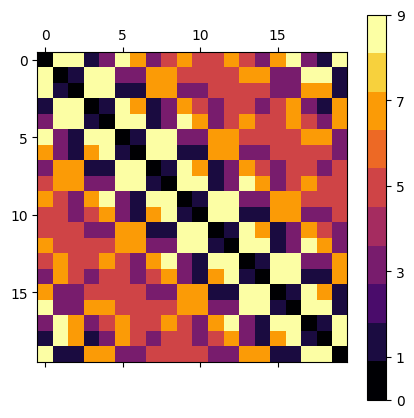}
        \caption{10-Mode MaxNTO Encoding}
        \label{fig:maxnto-ntos}
    \end{subfigure}
    \caption{Pairwise Non-trivial overlap.}{Pairwise non-trivial overlaps of operators in a) the Jordan-Wigner encoding with 4 modes. All values are one except for the diagonal, as each string necessarily has zero-NTO with itself. b) 10-Mode, 9-NTO MaxNTO encoding. Unlike TTs, the maximal NTO between pairs is equal to the number of fermionic modes minus one, with pairwise NTOs including all odd numbers.}
    \label{fig:nto}
\end{figure*}

Although the focus of this work is ternary tree encodings which are guaranteed to have 1-NTO between each pair of strings, we note that valid encodings exist for every odd-k NTO and these need not be constant among pairs of strings. The generalisation of the 3-NTO encoding presented by Miller \cite{miller_bonsai_2023} is straightforward given its representation in matrix-plot form. Figure \ref{fig:maxnto-ops} shows the operators of a 10-mode, 9-NTO encoding with Figure \ref{fig:maxnto-ntos} showing the number of NTOs between each pair of operators. In general this encoding has terms with $M-1$ non-trivial overlaps; for this reason we refer to it as the MaxNTO encoding. Unlike the Ternary Trees, the MaxNTO encoding has non-constant NTO between operators, and contains non-trivial overlaps with all odd numbers less than the number of modes, as can be seen in Figure \ref{fig:nto}.
\subsection{Ternary Tree Encodings}
\label{sec:tt}

Ternary trees (TTs) provide a method to produce a set of Pauli strings, which satisfy the criteria above for a Majorana-string encoding \cite{miller_bonsai_2023}.

\begin{figure}[h!]
    \centering
    \begin{subfigure}{0.5\textwidth}
        \includegraphics[width=0.9\linewidth, frame]{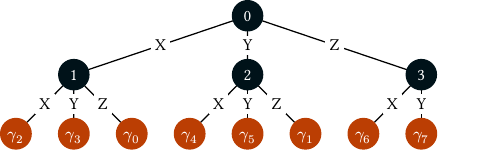}
        \caption{JKMN}
    \end{subfigure}\\
    \vspace{5mm}
    \begin{subfigure}{0.3\textwidth}
        \includegraphics[width=0.9\linewidth, frame]{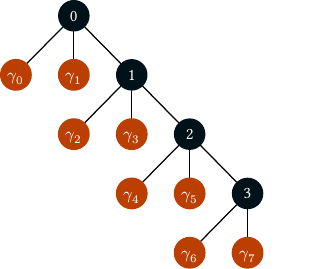}
        \caption{JW}
    \end{subfigure}
    \hfill
    \begin{subfigure}{0.3\textwidth}
        \includegraphics[width=0.9\linewidth, frame]{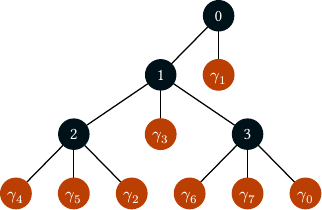}
        \caption{BK}
    \end{subfigure}
    \hfill
    \begin{subfigure}{0.3\textwidth}
        \includegraphics[width=0.9\linewidth, frame]{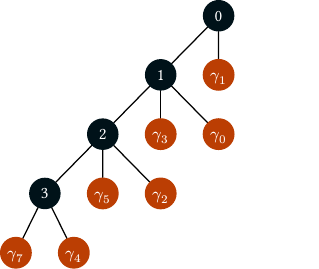}
        \caption{PE}
    \end{subfigure}
    \caption{Ternary tree structures of a) Jiang-Kalev-Mruczkiewicz-Neven (JKMN) b)Jordan-Wigner (JW) c) Bravyi-Kitaev (BK) d) Parity (PE) encodings \cite{jiang_optimal_2020} for four modes. Nodes are shown in black and leaves in red. Each node is enumerated with a qubit index, while each leaf has an associated Majorana operator $\gamma_i$. Edges between nodes in a) show the Pauli operator associated to the edge. By convention outward edges point downward and $\hat{X}$, $\hat{Y}$, $\hat{Z}$ are arranged as left, centre and right respectively.}
    \label{fig:standard-trees}
\end{figure}

For a system of $M$ fermionic modes, we define a graph $G(V,E)$ with vertices $V=\set{v}:|V|=M$ and edges $E=\set{e}: |E|=3M$. A qubit index is assigned to each node, which gives the position of operators associated to its outward edges.

Each node has three outward edges, one for each of the Pauli operators $\hat{X}$, $\hat{Y}$ and $\hat{Z}$. Each edge can connect to a child node, or be an unpaired "leaf". All nodes except the root node have a single inwards edge, one of the edges of its parent node. Going forward we refer to these with a combination of operator and parent/child/leaf. For example, a node could be said to have an X-Child, Z-Leaf or Y-Parent. 

Pauli-strings are constructed by following a path from each of the leaves to the root node, appending operators to the string according to each edge traversed \cite{miller_bonsai_2023}. Each path diverges from each of the others at exactly one node, resulting in a constant 1-NTO for each pair of operators. The number of leaves for a ternary tree will be $2M+1$, so by convention the leaf which is reached by only taking Pauli-Z edges from the root node is removed from the set. 

As an example, Figure \ref{fig:standard-trees} shows the tree structure for 4-mode JW, PE, BK and JKMN encodings. Black circles show the nodes of the tree, while red circles are the leaves.

Given their constant 1-NTO Majorana-operators, TTs represent a restricted class of Majorana-string encodings. Optimisation of TTs therefore cannot generally guarantee an optimal solution among all valid Majorana-string encodings. However, non-TT encodings can be reached by applying a unitary transformation to a TT \cite{chiew_ternary_2024, yu_clifford_2025}. An interesting avenue for future research is to determine which, or whether, TTs are an efficient initial state for unconstrained optimisation.

\section{Optimisation}
\label{sec:optimisation}

With any optimisation method, two things are needed. Firstly, well-motivated cost functions. Secondly, degrees of freedom over which to optimise. We consider each of these in turn within this section.

\subsection{Pauli-weight}
\label{sec:pauli-weight}
We define the Pauli-weight $W_P$ of a Pauli-string $S_{i}=(P_{0}^{i},\dots,P_{M}^{i})$ as the total number of qubits on which a non-identity operator is performed.
\begin{equation}
    W_{P}(S_{i}) = |P\ne \hat{I}|:P\in \set{S_{i}}.
\end{equation}
For a qubit-Hamiltonian composed of multiple strings, $\mathcal{H}_{q}=\sum_{i}h_{i}\equiv\sum_{i}c_{i}S_{i}$, the Pauli-weight is the sum of the weights of strings appearing in the Hamiltonian.
\begin{equation}
    W_{P}(\mathcal{H}_{q}) = \sum_{i}|P\ne \hat{I}|:P\in \set{S_{i}}.
\end{equation}
For quantum algorithms which rely on determining the expectation value of an operator, such as the VQE, the number of samples required for each term scales with its coefficient \cite{ralli_implementation_2021}. Reducing the Pauli-weight decreases the total gate and readout error, and allows for more measurements to be made in parallel. In the case of algorithms which implement fermionic operators as part of a circuit, the Pauli-weight of these operators should be kept low to minimise gate errors and circuit depth. 

\subsection{Coefficient-scaled Pauli-weight}
\label{sec:coeff-pauli-weight}
A related figure of interest is the coefficient-scaled Pauli-weight $W_{CP}$, which is defined for a qubit-Hamiltonian $\mathcal{H}_{q}=\sum_{i}h_{i}\equiv\sum_{i}c_{i}S_{i}$, where $c_{i}$ are complex scalar coefficients:

$$W_{CP}(h_{i}) = |c_{i}|\times|P\ne \hat{I}|:P\in \set{S_{i}}$$
$$W_{CP}(\mathcal{H}_{q}) = \sum_{i}|c_{i}|\times|P\ne \hat{I}|:P\in \set{S_{i}}$$

The coefficient-scaled Pauli-weight is relevant in the construction of circuits for the qDRIFT simulation algorithm \cite{li_huffman-code-based_2025}. As described in Section \ref{sec:qdrift}, terms in the Hamiltonian of interest are sampled at random according to the absolute value of their coefficients. Optimising an encoding such that the terms with the largest coefficients have the smallest Pauli-weight results in circuits with lower total depth for the same Hamiltonian terms \cite{li_huffman-code-based_2025}.

\subsection{Symmetries and Enumeration Schemes}
\label{sec:symmetry}

\begin{figure}[h!]
    \centering
    \begin{subfigure}[b]{0.3\textwidth}
        \includegraphics[width=0.9\linewidth, frame]{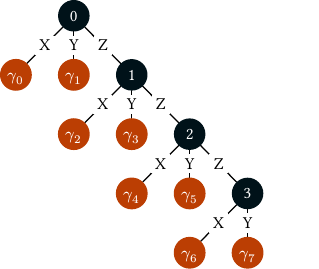}
    \end{subfigure}
    \begin{subfigure}[b]{0.3\textwidth}
        \includegraphics[width=0.9\linewidth, frame]{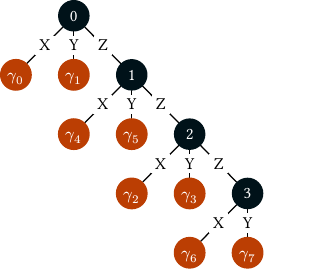}
    \end{subfigure}
    \begin{subfigure}[b]{0.3\textwidth}
        \includegraphics[width=0.9\linewidth, frame]{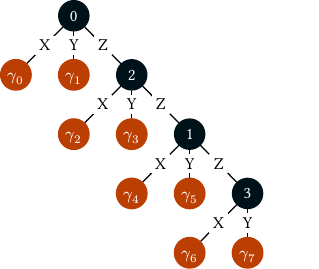}
    \end{subfigure}\\
    \begin{subfigure}[b]{0.2\textwidth}
        \includegraphics[width=\textwidth]{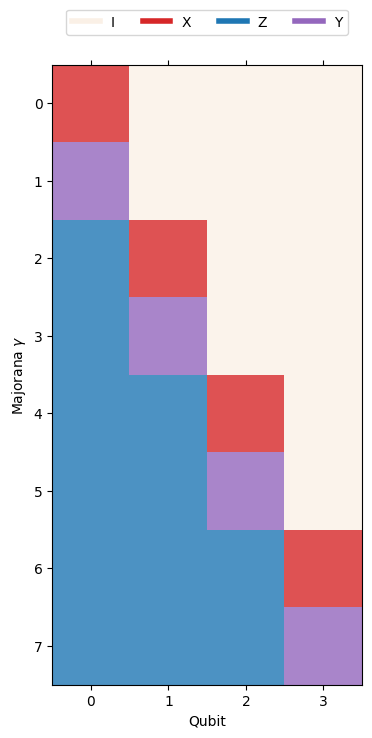}
    \caption{}
    \end{subfigure}
    \hspace{0.1\textwidth}
    \begin{subfigure}[b]{0.2\textwidth}
        \includegraphics[width=\textwidth]{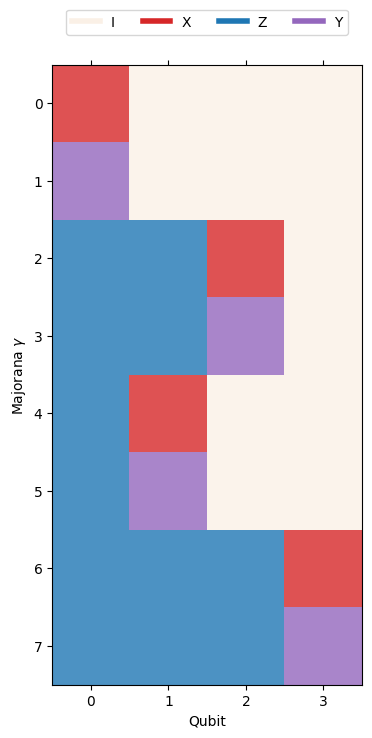}
    \caption{}
    \end{subfigure}
    \hspace{0.1\textwidth}
    \begin{subfigure}[b]{0.2\textwidth}
        \includegraphics[width=\textwidth]{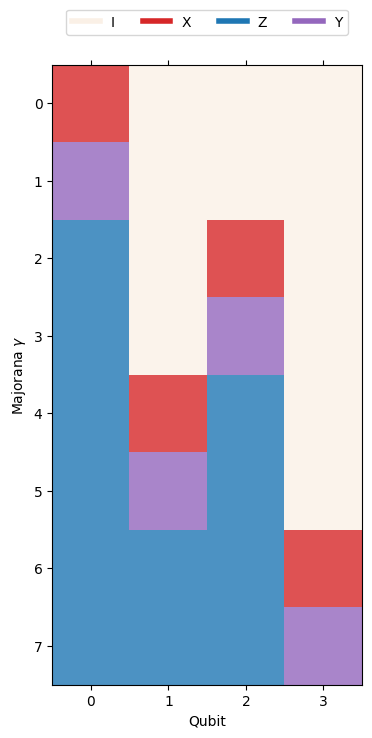}
        \caption{}
    \end{subfigure}
    \caption{Enumerations of the four mode Jordan-Wigner encoding. a) The naive enumeration, in which fermionic mode index and qubit index are equal, and increasing with distance from the root node. b) An altered fermionic mode enumeration, in which the Majorana-operators assigned to fermionic modes $1$ and $2$ have been swapped. c) An altered qubit enumeration, in which the indices of qubits $1$ and $2$ have been swapped.}
    \label{fig:enumerations}
\end{figure}

Optimisation of fermion encodings can be thought of as comprising two separate strategies, structuring of the initial encoding and permutation of symmetries within the encoding. For the case of Majorana-string encodings, the first step consists of defining a set of $2M$ Pauli-strings $S_i=(P_{0}^i,\dots,P_{M}^i)$ satisfying the conditions to form an encoding.

The second step consists of applying an \textit{enumeration scheme} to the strings $S \in \set{S}$. Pairs of strings are assigned to pairs of Majorana-operators $\gamma\in\set{\gamma}$ in the Hamiltonian $(S_{2i}, S_{2i+1}) \to (\gamma_{\alpha}, \gamma_{\beta})$, and positions in the string are assigned to physical qubits $P^{i}_{j} \to \hat{P}_{q \in \set{Q}}$. Figure \ref{fig:enumerations} shows the effect of changes to the enumeration scheme on operators of the 4-mode Jordan-Wigner encoding.

\begin{figure}[h!]
    \centering
    \includegraphics[width=0.5\linewidth]{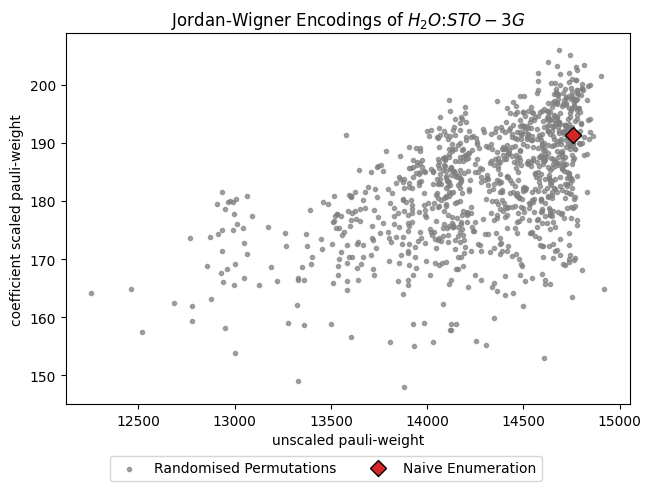}
    \caption{Pauli-weight and coefficient-scaled Pauli-weight of 1000 random enumerations of fermionic modes for the water molecule in an \emph{STO-3G} basis, generated using the `ferrmion' software package.[citation anonymised]}
    \label{fig:unoptimised-jw}
\end{figure}

As each fermionic operator in the Hamiltonian is associated to a coefficient from the one or two electron integrals, by strategically assigning strings to Majorana operators the qubit Hamiltonian can be constructed such that terms which have a zero coefficient can be assigned to those with the greatest Pauli-weight. Further those with the coefficients of greatest amplitude can be assigned to the operators with the lowest Pauli weight. The practical consequences of this are demonstrated in Figure \ref{fig:unoptimised-jw}, which shows the distribution of $W_{P}(\mathcal{H}_{\gamma})$ and $W_{CP}(\mathcal{H}_{\gamma})$ for 1000 random enumerations of Majorana-operator pairs, for the water molecule in an \emph{STO-3G} basis. Shown also is the naive enumeration, the standard form of Jordan-Wigner encoding \cite{chien_optimizing_2022}.

The enumeration of qubit indices has a less straightforward relationship with the resulting circuit. Consideration must be made of the device topology and native gate set in addition to the Hamiltonian terms \cite{miller_bonsai_2023, miller_treespilation_2024}. Compilation and qubit-routing techniques are able to combine operations or remove redundant sequences. As a result, to accurately assess the effect of qubit enumeration, full circuits should be compiled. 
The method we now present is designed to optimise mode enumeration for arbitrary tree structures, and thus can be applied to encodings which are derived from device topology by, for instance, the Bonsai Algorithm \cite{miller_bonsai_2023}.
The optimisation of qubit enumeration is a promising area for future research.

\section{TOPP-HATT}
\label{sec:TOPP-HATT}

The structure of our method is shown in Algorithm \ref{alg:TOPP-HATT}. Required inputs are a Majorana-Hamiltonian $\mathcal{H}_{\gamma}$ for the system of interest, and a TT graph $G(V,E)$. Some initial setup is required before an optimised enumeration scheme is built iteratively.

\subsection{Setup}
We begin by building a tree which represents the \emph{naive} enumeration. To do so, a set of nodes is created, one for each of the $M$ fermionic modes of the Hamiltonian. For ternary trees, a single qubit is required for each mode, so each node is assigned a qubit index $q_i\in \mathcal{Q} \equiv \{q_0,...,q_M\}$. Each node has a \emph{pair} of leaves $L_i$ assigned to its $X$ and $Y$ edges. These leaves are associated to a pair of Majorana operators, labelled according to Equations \ref{eq:annihiliation} and \ref{eq:creation}, i.e. $L_i\to (\gamma_{2i},\gamma_{2i+1})$. The naive tree is built iteratively by attaching new child nodes to existing nodes, according to the graph $G(V,E)$, and beginning with the root node. Any leaf which is replaced by a child node is moved to the unoccupied $Z$ edge of the child node. As we shall discuss in Section \ref{sec:vacuum}, this ensures vacuum preservation. Ultimately, the naive tree describes the position of each node and leaf, with their indices corresponding to the naive enumeration of Majorana operators. Alternative mode enumerations of the encoding are given by permutations of assignments of each mode's $f_j\in \set{f}$ associated Majorana-operator indices to a pair of leaves $f_j: (\gamma_{2j},\gamma_{2j+1}) \to L_k$. It is these indices which we optimise over in the following method.

To proceed with optimisation, the naive tree is used to initialise a set of restrictions on the possible choice of leaf indices for each node. These are explained in detail in the following section. 
The naive tree also makes it possible to define a map from the location of a leaf in the naive tree $(q\in \mathcal{Q}, Edge\in\set{X,Y,Z})$ to the location of its pair. This is used to ensure that when one of a pair of leaves is assigned a Majorana-operator, $\gamma_{2m}(\gamma_{2m+1})$, the other leaf in the pair can be updated with the corresponding operator $\gamma_{2m+1}(\gamma_{2m})$ in $\mathcal{O}(1)$ time.

\subsection{Restrictions}
Given that we require optimisation does not change the underlying graph structure of an input tree, we are restricted to optimisation only over the indices of leaves assigned to each node. In the following section we detail the procedure for defining these restrictions such that we can guarantee: (i) the encoding is a valid Majorana-string encoding according to the criteria in Section \ref{sec:mse}, (ii) the encoding topology is preserved, and (iii) the encoding is vacuum preserving.

\subsubsection{Operator Independence}
\label{sec:restrictions}
The simplest restriction, which is placed on all TTs by convention, is to remove the Majorana-operator which results from taking only Z-edges from the root node outward. This ensures that the criteria for valid Majorana-string encoding are maintained. The naive tree therefore does not have a leaf at this position and no assignment can be made to this edge during optimisation.

\begin{figure}[h!]
    \centering
    \includegraphics[width=0.3\linewidth]{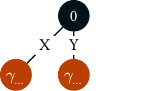}
    \caption{Single node ternary tree}
    \label{fig:single-node}
\end{figure}

The simplest possible tree is the single-node tree, shown in Figure \ref{fig:single-node}, which has two leaves corresponding to two Majorana operators $\set{\gamma} = \set{\gamma_0,\gamma_1}$. The All-Z-Leaf restriction applies so no Z-leaf is present. If the Z-leaf were present, it would be possible to generate any one of the three Majorana operators $\{X,Y,Z\}$, from the other two, thus violating criterion 4 of Section \ref{sec:encoding}.

\subsubsection{Tree Structure}
Given that our aim is to preserve the structure of an input tree, all nodes with one or more child nodes must retain those children on the same edge. Additionally, the positions of each qubit index must not change. For TTs derived from the connectivity graph of a QPU, retaining qubit indices ensures operators are mapped to specific physical qubits.

\begin{figure}[h!]
    \centering
    \includegraphics[width=0.5\linewidth]{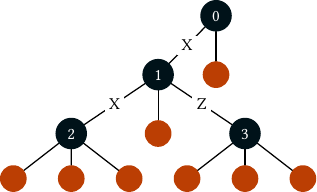}
    \caption{The nodes of the four-mode Bravyi-Kitaev encoding must retain their positions and qubit indices.}
    \label{fig:bk-nodes}
\end{figure}

As an example, Figure \ref{fig:bk-nodes} shows the four-mode Bravyi-Kitaev tree. During optimisation, the nodes must have positions and qubit indices as follows: $root\to0$, $X\to1$, $XX\to2$, $XZ\to3$.

\subsubsection{Vacuum Preservation}
\label{sec:vacuum}

To ensure that the fermionic vacuum state $\ket{}_{f}$ is encoded as the qubit state $\ket{0}_{q}^{N}$, we are restricted in the way Majorana-operators can be paired. 

\begin{figure}[ht]
    \centering
    \begin{subfigure}{0.9\textwidth}
        \includegraphics[width=0.3\linewidth]{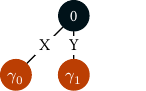}
        \includegraphics[width=0.3\linewidth]{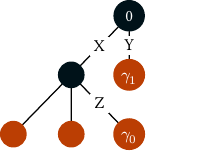}
        \includegraphics[width=0.3\linewidth]{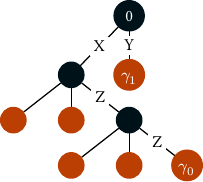}
        \caption{Replacing leaves}
    \end{subfigure}\\
    \begin{subfigure}{0.5\textwidth}
        \includegraphics[width=\linewidth]{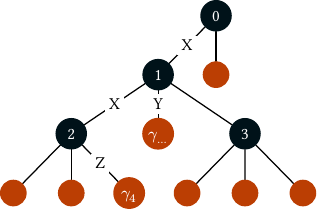}
        \caption{Updating pair restrictions}
    \end{subfigure}
    \caption{a) Vacuum preservation can be guaranteed by moving a replaced leaf to the Z-Edge of the node which is replacing it. b) When the Z-Leaf of node 2 is assigned an optimal value, we can determine the required value of the Z-Leaf of node 3. In this case $\gamma_4$ has been assigned to the leaf at $XXZ$, the pair of which is at $XY$. Given the definition of fermionic operators in \ref{eq:annihiliation} and \ref{eq:creation}, the pair of $\gamma_4$ must be $\gamma_5$. So we assign both leaves in the pair as follows: $L\equiv\{XXZ, XY\}\to (\gamma_4, \gamma_5)$. }
    \label{fig:vacuum-preservation}
\end{figure}

For a single node, as in Figure \ref{fig:single-node}, it begins with a valid pair on edges $X$ and $Y$. If we wish to add a child node to one of these edges, (for instance, the X-Edge), we must find a new position for the replaced leaf. We can ensure vacuum preservation by always re-assigning a replaced leaf to the Z-Edge of the node which replaces it. See Figure \ref{fig:vacuum-preservation} for an illustrated example. Given that the naive tree is vacuum preserving, to retain this property the set of pairs of leaves $\{L\}$ must not change during optimisation.

\subsubsection{Example Initial Restrictions}
To make the above concrete, we take the four-mode Bravyi-Kitaev tree of Figure \ref{fig:bk-nodes} as an example. Table \ref{tab:res-bk} gives the initial restrictions on this tree, while Table \ref{tab:pair-bk} gives the map between its pairs of leaves.

\begin{table}[h!]
    \centering
    \begin{tabular}{|c|c|c|c|}
    \hline
    \hline
        Node & X-edge & Y-edge & Z-edge \\
        \hline
         0&Node(1) & OddLeaf & Empty\\
         1&Node(2) & OddLeaf & Node(3)\\
         2&EvenLeaf & OddLeaf & EvenLeaf\\
         3&EvenLeaf & OddLeaf & EvenLeaf\\
        \hline
        \hline
    \end{tabular}
    \caption{Initial restrictions on the 4-mode Bravyi-Kitaev tree of Figure \ref{fig:bk-nodes}.}
    \label{tab:res-bk}
\end{table}

\begin{table}[h!]
    \centering
    \begin{tabular}{|c|c|c|}
    \hline
    \hline
        Pair & Even Parity Leaf & Odd Parity Leaf \\
        \hline
        $L_0$ & (3,Z) & (0,Y) \\
        $L_1$ & (2,Z) & (1,Y) \\
        $L_2$ & (2,X) & (2,Y) \\
        $L_3$ & (3,X) & (3,Y) \\
        \hline
        \hline
    \end{tabular}
    \caption{Leaf pairs of the 4-mode Bravyi-Kitaev tree of Figure \ref{fig:bk-nodes}. When one leaf in a pair is assigned an index, we also update the index of its pair, ensuring the encoding is vacuum preserving.}
    \label{tab:pair-bk}
\end{table}

\subsection{Iteration Loop}

Before the iterative optimisation procedure can begin, a set of \emph{unassigned} mode indices is created, and populated by fermionic modes for which Majorana operators have not been assigned to leaves (initially, all of them). During optimisation we will draw indices from this set, assigning to leaves the Majorana operator indices which correspond to unassigned modes. We initialise also a set of \emph{active} nodes. In contrast to the Huffman TT or HATT, we restrict our search to nodes which:

\begin{enumerate}
    \item Have no children with unassigned leaves;
    \item Are at the maximum distance from the root node of all nodes which meet criterion 1.
\end{enumerate}

\begin{figure}[ht]
    \centering
    \includegraphics[width=0.5\linewidth]{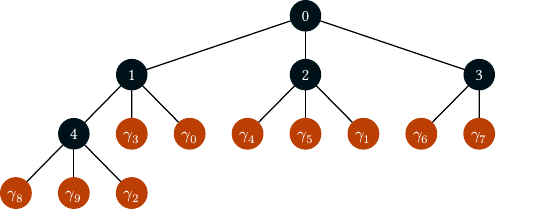}
    \caption{Five-mode JKMN encoding. During the first iteration of TOPP-HATT, although nodes 2,3 and 4 have no child nodes, only node 4 is at the maximum distance from the root node. Therefore the active nodes of this graph are initialised as $active_0 =\{4\}$. In the subsequent iteration $active_1=\{1,2,3\}$.}
    \label{fig:jkmn-five}
\end{figure}

For linear encodings such as Jordan-Wigner and Parity, only one node will be active at a time. In general, however, multiple eligible nodes can share the maximum distance from the root node, in which case all of these are active. The Bravyi-Kitaev encoding in Figure \ref{fig:standard-trees} for example, during the first iteration, has two nodes at a distance of 2 from the root with no child nodes, $\{2,3\}$, so these are both \emph{active}. Optimisation begins with the left-most node, $2$ before proceeding to subsequent nodes (here $3$). In contrast, although the 5-mode JKMN encoding of Figure \ref{fig:jkmn-five} has three nodes without children $\{4,2,3\}$, only one of them ($4$) is at the maximum root-distance of 2. In the first iteration node $4$ is therefore guaranteed to be assigned leaf indices, with its parent node, $1$, added to the active nodes of the second iteration $\{1,2,3\}$.

Leaf indices are assigned for each of the $M$ graph nodes iteratively. At the beginning of each iteration, the minimum weight, $min$, qubit index of the parent for which the minimum was found, $minparent$, and $selection$ of indices are reset. For each active node, $A$, we determine the set of possible assignments to each of its outward edges $\set{x}, \set{y}, \set{z}$. During the first iteration, all outward edges are necessarily leaves, as the active nodes are terminal nodes of branches, but during subsequent iterations nodes may have children that they are required to retain. In such cases, the only possible value for this edge is the child node. 

For each possible value, including leaves, nodes and the empty all-z branch, we require a unique identifier. Possible leaf indices are drawn from the subset of \emph{unassigned} modes for which leaves are still to be assigned, with the index parity determined by the set of restrictions $R$. We follow the convention of Liu et al. \cite{liu_hatt_2025} by setting the index of the all-z branch to $N_{leaves}+1$, and additionally index nodes according to their qubit index $q_i\to q_i+N_{leaves}$.

Taking the Cartesian product of possible indices, the total Pauli-weight is calculated for each combination. The Pauli-weight for a single term $h_{i}$ of the Majorana-encoded Hamiltonian $\mathcal{H}_{\gamma}$ is found by replacing occurrences of indices $x$, $y$ and $z$ with their respective Pauli-operators ($\hat{X},\hat{Y},\hat{Z}$), and any other indices with the identity operator $\hat{I}$ \cite{liu_hatt_2025}. We do not consider the sign or magnitude of Hamiltonian coefficients, only the presence or absence of a Pauli-operator. Note that the index of the all-z branch will not appear in the Hamiltonian as no Majorana operator is assigned to it. Reducing the resulting string with identities for Pauli-operators results in either a single Pauli-operator or the identity, thus each term has a weight of 0 or 1 for each node. The total Pauli-weight of the combination $(x,y,z)$ is the sum of that for individual terms, $W^{(x,y,z)}_{P}(\mathcal{H}_\gamma)\le |\{h\}|$. 

\begin{algorithm}[h!]
    \caption{TOPP-HATT}
    \label{alg:TOPP-HATT}
    \begin{algorithmic}[1]
        \Procedure{TOPP-HATT}{$\mathcal{H}_{\gamma}, G(V,E)$}
        \State $tree \gets NaiveTernaryTree(G(V,E))$
        \State $R\gets InitialiseRestrictions(tree)$
        \State $L \gets InitialiseLeafPairMap(tree)$
        \State $unassigned \gets \{0..M\}$
        \State $active\gets InitialiseActiveNodes(R)$
            \For{$i=1,\dots,M$}
                \State $min \gets \infty$
                \State $parent \gets Null$
                \State $selection \gets (Null, Null. Null)$
                \For {$A$ in $active$}
                    \State $(\set{x}, \set{y},\set{z}) \gets AllowedIndices(R,A)$
                    \For{$(x,y,z)$ in $CartesianProduct(\set{x}, \set{y}, \set{z})$}
                        \State weight $\gets W_P(\mathcal{H}_{\gamma},x,y,z) $
                        \If{$weight < min$}
                            \State $min \gets weight$
                            \State $selection \gets (x,y,z)$
                            \State $minparent \gets A$
                        \EndIf
                    \EndFor
                \EndFor
                \State $active\gets UpdateActiveNodes(assigned,\ parent)$
                \State $R \gets UpdateRestrictions(R,\ parent,\ z)$
                \State $tree \gets AssignLeaves(minparent, x,y,z)$
                \State $tree \gets AssignPair(L(z))$
                \State $unassigned \gets UpdateUnassignedModes(x,y,z)$
                \State $\mathcal{H}_{\gamma}\gets ReduceHamiltonian(\mathcal{H}_{\gamma},\ parent,\ selection)$
            \EndFor
        \EndProcedure
    \end{algorithmic}
\end{algorithm}

At this point we update the tree, using the optimal \emph{selection} obtained. Additionally if the $Z$-edge is a leaf, we assign a Majorana operator to its pair, as described in Section \ref{sec:vacuum}. To ensure the value of this leaf is retained in future iterations, we update the set of restrictions, disallowing any other choice. Modes for which leaves have been assigned are then removed from the \emph{unassigned} set.

Finally, the Hamiltonian is reduced, using the method described by Liu et al. \cite{liu_hatt_2025}. Given that some node $A$ is assigned the selection $(x,y,z)$, indices in the selection will be mapped to Majorana operators which act identically on nodes appearing between the root and $A$: $\gamma_x \equiv \gamma_y \equiv \gamma_z$ $\forall q\in Path(root\to A)$. For the remainder of iterations of the procedure, wherever these Majorana operators appear in $H_{\gamma}$, they will act with the same Pauli-operator. Therefore, it is possible to simplify $H_{\gamma}$ by substituting an index for the assigned node in place of each of $x,y,z$. For example, substituting $x$ and $y$ in the following: $h_i(x,k,j,y)\to h_i(A,k,j,A)$. This term can be simplified further by removing pairs of duplicate indices, $h_i(A,k,j,A)\to h_i(k,j)$.

\section{Results}
\label{sec:results}
This method, and all those it is compared against, have been implemented in the \texttt{ferrmion} software package [citation anonymised]. All source code and input data are available on our \href{https://github.com/UCL-CCS/ferrmion}{GitHub Repository} \emph{https://github.com/UCL-CCS/ferrmion}, together with an interactive notebook to reproduce the below results. qDRIFT circuits were compiled using the \texttt{TN4QA} software package, \cite{mingare_tn4qa_2025}, and transpiled using the \texttt{Qiskit} (2.3.0) transpiler \cite{kremer_practical_2025}. Runtime measurements were obtained using a single core of the Apple M3 Pro chip. TOPP-HATT code was compiled with rustc 1.95.0-nightly (47611e160 2026-02-12).

In the following section we present distributions of encodings of the water molecule in the minimal \emph{STO-3G} basis. The naive enumeration is shown together with results obtained by simulated annealing of the Majorana-operator enumeration, using $W_P$ and $W_{CP}$ as the cost function. 

\begin{figure}[h!]
    \centering
    \begin{subfigure}[b]{0.45\textwidth}
        \includegraphics[width=\textwidth]{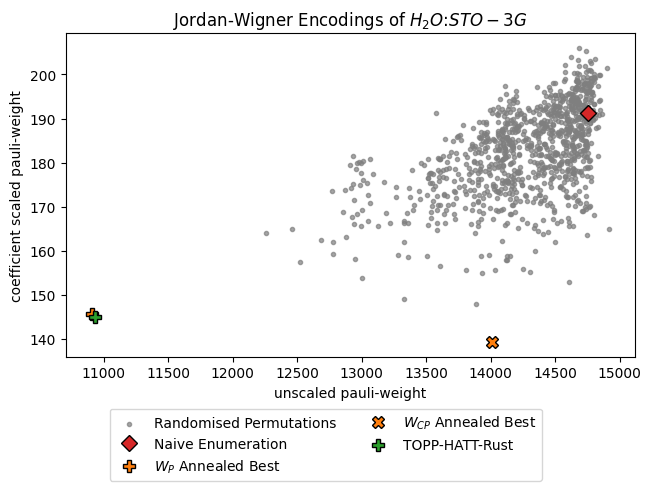}
    \end{subfigure}
    \begin{subfigure}[b]{0.45\textwidth}
        \includegraphics[width=\textwidth]{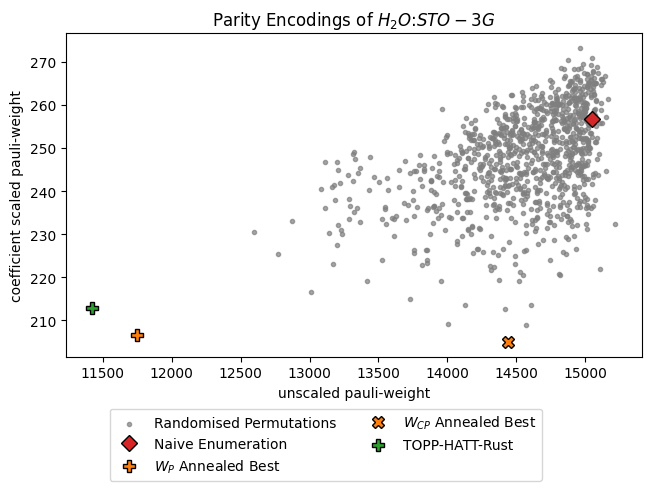}
    \end{subfigure}
    \\
    \begin{subfigure}[b]{0.45\textwidth}
        \includegraphics[width=\textwidth]{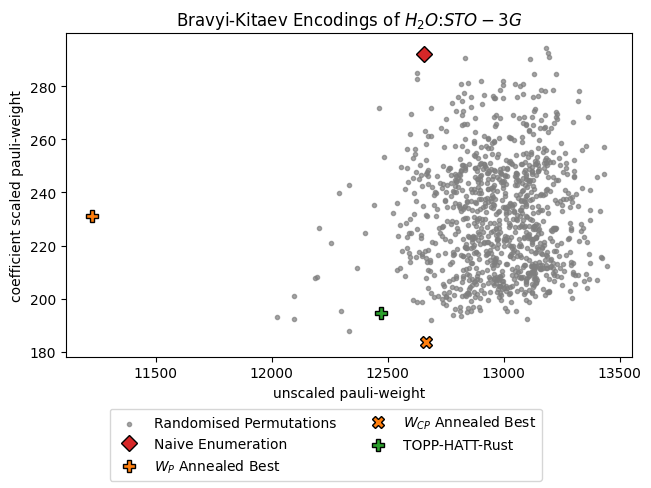}
    \end{subfigure}
    \begin{subfigure}[b]{0.45\textwidth}
        \includegraphics[width=\textwidth]{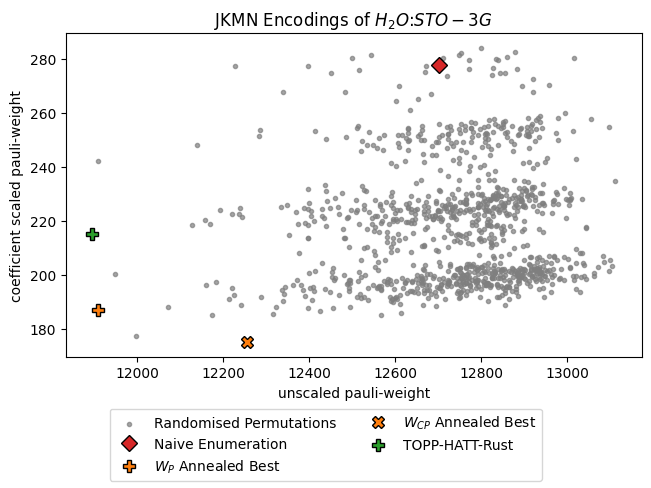}
    \end{subfigure}
    \caption{Permutations of the Jordan-Wigner (top-left), Parity (top-right), Bravyi-Kitaev (bottom-left) and JKMN (bottom-right) encodings for $H_2O:STO-3G$ (14 Modes). On the \emph{x}-axis of each plot is the average Pauli-weight of terms in the encoded Electronic Structure Hamiltonian, while the \emph{y}-axis is the average coefficient-scaled Pauli-weight. Each plot shows 1000 random enumerations of the modes in grey, the naive enumeration as a red diamond, the simulated-annealing optimised enumeration as an orange circle and the TOPP-HATT result as a green cross.}
    \label{fig:TOPP-HATT-standard}
\end{figure}

We demonstrate runtime scaling of our method using a selection of molecules for both the \emph{STO-3G} and \emph{6-31G} basis sets to make clear that our method is generally applicable, including beyond the current limitations of available quantum processors. 

\subsection{Standard Encodings}

To demonstrate our method, we first show results for standard encodings with 14 modes. Figure \ref{fig:TOPP-HATT-standard} shows results for each of Jordan-Wigner, \cite{jordan_uber_1928} Parity,\cite{bravyi_fermionic_2002} Bravyi-Kitaev, \cite{bravyi_fermionic_2002} and JKMN \cite{jiang_optimal_2020}.

Although each of the standard encodings serves as a well-defined benchmark, the Jordan-Wigner encoding is in widespread use in quantum simulation of chemistry. In particular, the Local Unitary Cluster Jastrow ansatz employed in Quantum Selected Configuration Interaction methods is reliant upon the association of fermionic mode occupation to qubit spin which JW provides \cite{JastrowtypeDecompositionQuantum2020}. 

Our results show that TOPP-HATT is particularly effective for the linear Jordan-Wigner and Parity encodings, with reduced benefit for the binary Bravyi-Kitaev and ternary JKMN trees. The method as presented may result in sub-optimal enumerations as we assign the pair of leaves in the $selection$ without consideration of its impact on Pauli-weight. Leaf assignments only consider the Pauli-weight contribution of a single node in each iteration and do not account for interactions between possible assignments on different active nodes. In the case of the Jordan-Wigner encoding, all pairs of leaves share a single parent node, so no such interactions are possible, and enumerations are therefore guaranteed to be globally optimal. 

\subsection{Hamiltonian-optimised Encodings}

\begin{figure}[h!]
    \centering
    \begin{subfigure}[b]{0.35\textwidth}
        \includegraphics[width=\textwidth]{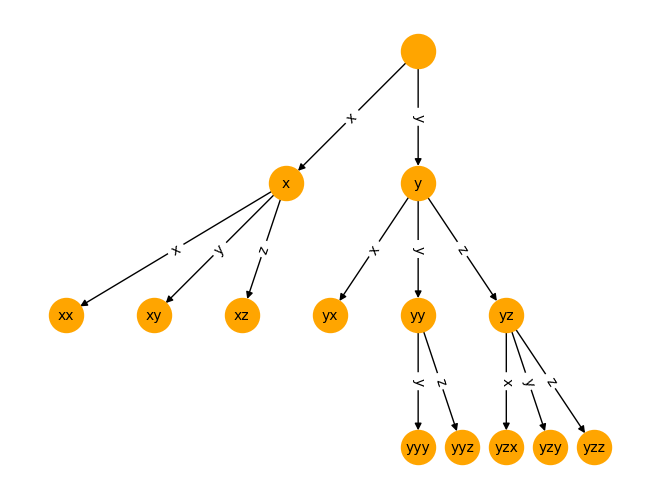}
    \end{subfigure}
    \begin{subfigure}[b]{0.55\textwidth}
        \includegraphics[width=\textwidth]{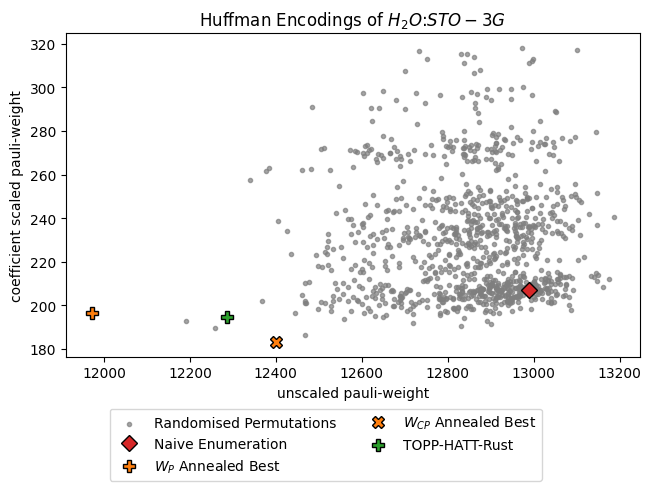}
    \end{subfigure}
    \\
    \begin{subfigure}[b]{0.35\textwidth}
        \includegraphics[width=\textwidth]{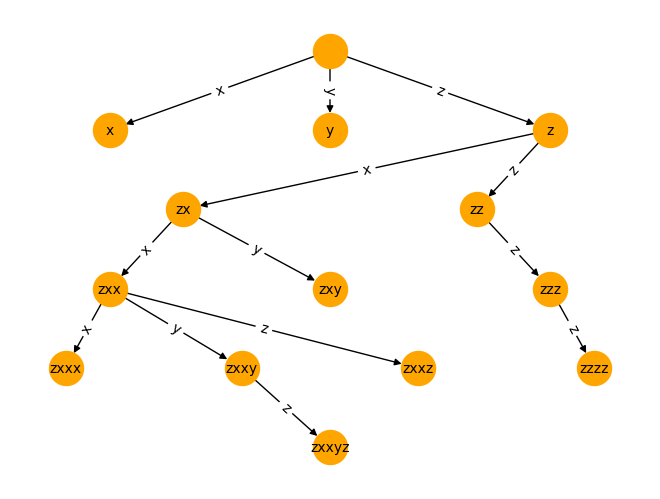}
    \end{subfigure}
    \begin{subfigure}[b]{0.55\textwidth}
        \includegraphics[width=\textwidth]{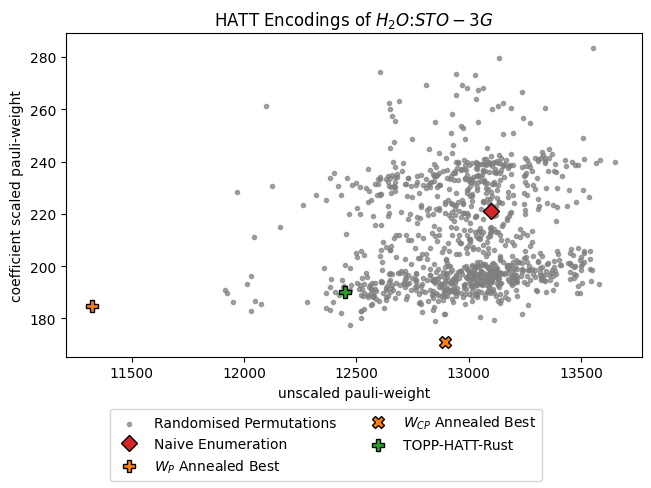}
    \end{subfigure}
    \caption{Permutations of the Huffman-code Ternary Tree (top row) and Hamiltonian-Adaptive Ternary Tree (bottom row) for $H_2O:STO-3G$ (14 Modes). On the x-axis of each plot is the average Pauli-weight of terms in the encoded electronic structure Hamiltonian, while the y-axis is the average coefficient-scaled Pauli-weight. Each plot shows 1000 random enumerations of the modes in grey, the naive enumeration as a red diamond (which in the case of Huffman and HATT encodings are optimised), the simulated-annealing optimised enumeration as an orange circle and the TOPP-HATT result as a green cross.}
    \label{fig:TOPP-HATT-ham-optimised}
\end{figure}

We now compare the results of our method to methods which are optimised according to the Hamiltonian only, namely Huffman-code TT \cite{li_huffman-code-based_2025} and Hamiltonian-adaptive TT \cite{liu_hatt_2025}. In each case, the Hamiltonian-optimised method is first run, resulting in a TT with some unconstrained structure. This tree structure is then given as input to TOPP-HATT.

In the case of the Huffman-code TT, which is designed to minimise the coefficient-scaled Pauli-weight, our method shows slightly improved performance on this metric, while reducing the Pauli-weight.

The slight difference in our method to the results obtained by HATT is due to differing heuristics for the ordering of nodes which are optimised and the choice between selections of equal weight. Regardless, our method performs comparably, as expected.

\subsection{Device-optimised Encodings}
\begin{figure}[h!]
    \centering\includegraphics[width=0.5\linewidth,frame]{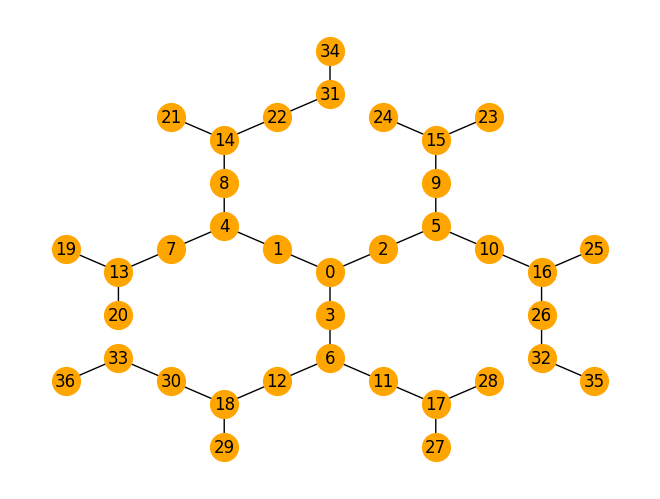}
    \caption{Qubit connectivity graph for a heavy-hex device, similar to that presented in Miller et al. 2023 \cite{miller_bonsai_2023}.}
    \label{fig:heavy-hex}
\end{figure}

The key benefit of our method is that it allows for the optimisation of TTs which are subgraphs of a given QPU connectivity graph. This reduces the requirement for SWAP gates by ensuring that Majorana-operators are formed of operators on contiguous qubits.

\begin{figure}[h!]
    \centering
    \begin{subfigure}[b]{0.35\textwidth}
        \includegraphics[width=\textwidth]{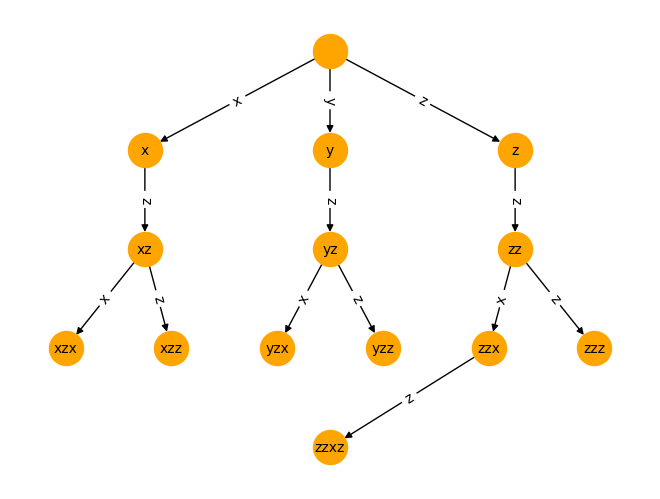}
    \end{subfigure}
    \begin{subfigure}[b]{0.55\textwidth}
        \includegraphics[width=\textwidth]{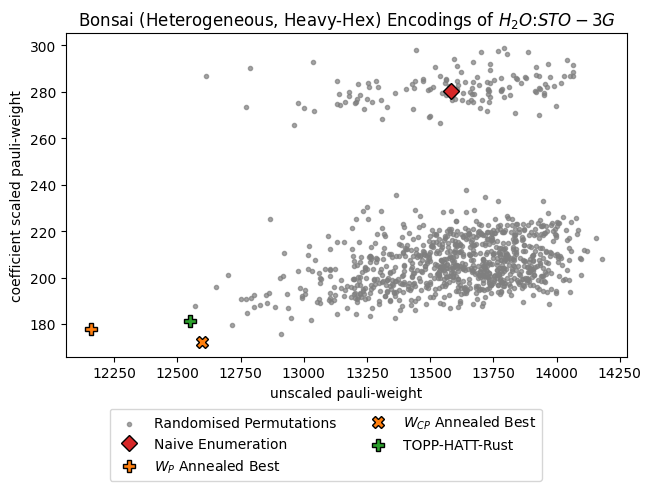}
    \end{subfigure}
    \\
    \begin{subfigure}[b]{0.35\textwidth}
        \includegraphics[width=\textwidth]{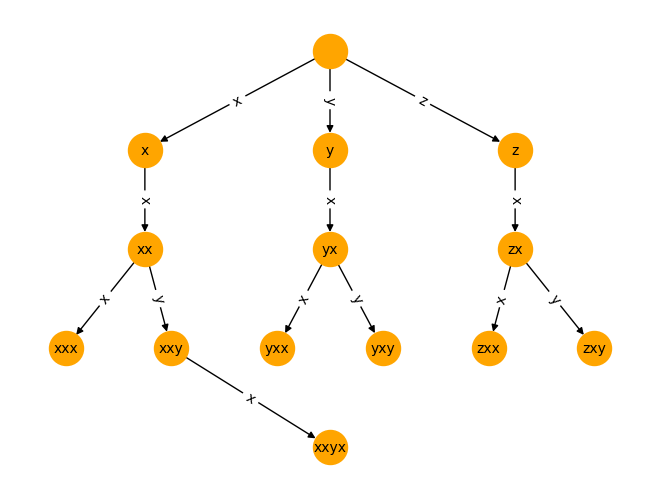}
    \end{subfigure}
    \begin{subfigure}[b]{0.55\textwidth}
        \includegraphics[width=\textwidth]{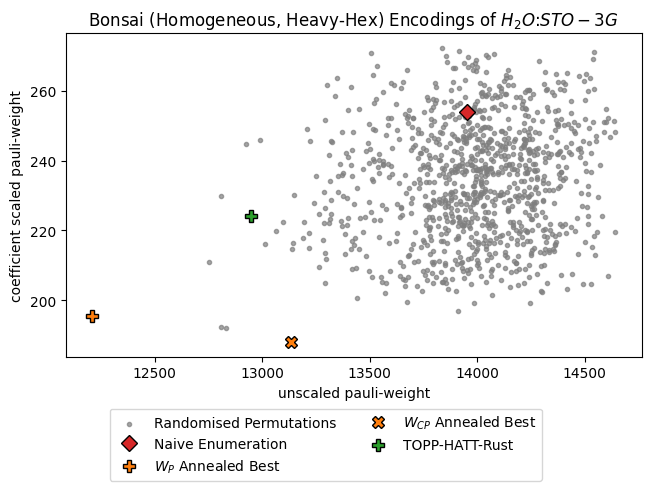}
    \end{subfigure}
    \caption{Permutations of the Heterogenous Bonsai Algorithm Ternary Tree (top row) and Homogeneous Bonsai Algorithm Ternary Tree (bottom row) for $H_2O:STO-3G$ (14 Modes). On the x-axis of each plot is the average Pauli-weight of terms in the encoded electronic structure Hamiltonian, while the y-axis is the average coefficient-scaled Pauli-weight. Each plot shows 1000 random enumerations of the modes in grey, the naive enumeration as a red diamond, the simulated-annealing optimised enumeration as an orange circle and the TOPP-HATT result as a green cross.}
    \label{fig:TOPP-HATT-bonsai}
\end{figure}

We demonstrate this by pairing our method with the Bonsai algorithm, which constructs minimal-depth trees from the connectivity graph of a device. Figure \ref{fig:heavy-hex} shows a connectivity graph for a 36-qubit device in heavy-hex layout, given as an example for the Bonsai algorithm \cite{miller_bonsai_2023}. Using this graph we construct a tree for each of the \emph{Heterogeneous} and \emph{Homogeneous} heuristics of the Bonsai algorithm.

\subsection{qDRIFT Circuit Depth}
\label{sec:qdrift}
The stochastic circuit compilation method of Campbell \cite{campbell_random_2019}, commonly known as the `qDRIFT' method provides an example of where the TOPP-HATT optimisation procedure can be used to assist near-term quantum algorithms for time-evolved Hamiltonian simulation. When implementing the time-evolution operator $U(t) = \mathrm{exp}\left(-iHt\right)$ for a given Hamiltonian, quantum circuits can be constructed deterministically via Trotter decompositions \cite{nielsen_quantum_2012}. However, implementing the full circuit requires a number of operators which scales polynomially with the total number of terms in the Hamiltonian. This quickly produces circuits of prohibitive depth for near-term quantum devices, even for Hamiltonians of modest size. The qDRIFT method addresses this issue by constructing circuits from only a limited set exponentiated Pauli terms, randomly sampled from the Hamiltonian according to their coefficient. For a qubit Hamiltonian of the form $\mathcal{H}_{q} = \sum_{j}c_{j}S_{j}$, as introduced in Section \ref{sec:pauli-weight}, the sampling probability of each term in the Hamiltonian is defined as $p_{j} = |c_{j}| / \lambda$, with $\lambda = \sum_{j} |c_{j}|$. For each sampled Pauli string $S_{j}^{s} \in S$ from a budget of $N_{s}$ samples, the unitary operator $U_{j}(t) = \mathrm{exp} \left( -i\lambda t \, \mathrm{sgn}(c_{j}) S_{j}^{s} / N_{s} \right)$ is constructed in the qDRIFT circuit. This circuit approximates the full evolution operator $U$.

The number of samples $N_{s}$ can be chosen according to a target precision $\epsilon$, with $N_{s} = \mathrm{ceil}\left( 2\lambda^{2}t^{2}/\epsilon \right)$, meaning that for a given precision and evolution time the circuit depth scales with $\lambda$, which is equivalent to the coefficient-scaled Pauli weight as discussed in Section \ref{sec:coeff-pauli-weight}. While TOPP-HATT optimisation is designed to reduce the unscaled Pauli-weight for a given tree topology, a practical consequence of this is a reduction in the coefficient-scaled Pauli-weight. We therefore use it as a pre-processing step on a Hamiltonian to reduce qDRIFT circuit depth. Below, we investigate this behaviour for untranspiled (raw) and transpiled (optimised for a specific quantum device) qDRIFT circuits.

\begin{figure}[h!]
    \centering
    \includegraphics[width=0.5\linewidth]{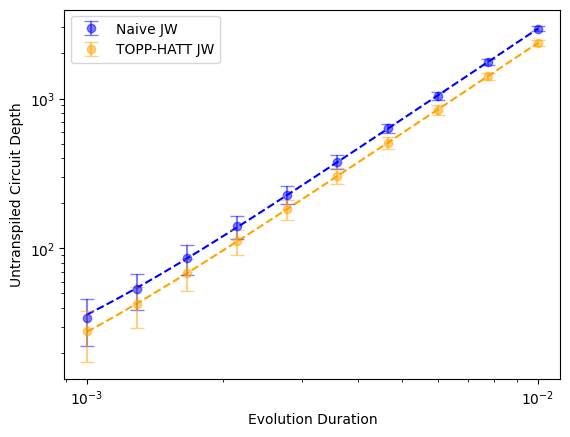}
    \caption{Untranspiled qDRIFT circuit depth in log scale as a function of evolution duration for water in the STO-3G basis set. Two encodings are used: naive Jordan-Wigner (shown in blue), and TOPP-HATT optimised Jordan-Wigner (shown in orange). The markers show the mean circuit depth from a batch of 100 qDRIFT circuits, and the errorbars are one standard deviation. The dashed lines are a fit to $k*depth^{2}$, with k(\small{TOPP-HATT})=$2.35e7$ and k(naive)=$2.93e7$.}
    \label{fig:qdrift_depths_water}
\end{figure}

Figure \ref{fig:qdrift_depths_water} shows the relationship between untranspiled circuit depth and evolution duration for a qDRIFT simulation of water in the \emph{STO-3G} basis, using the Jordan-Wigner (JW) encoding. qDRIFT circuits were constructed using the \texttt{TN4QA} software package \cite{mingare_tn4qa_2025}. 100 qDRIFT circuits were constructed at each evolution duration. The naive JW encoding is shown in blue, which represents the circuit depth scaling that could be expected without any consideration to the fermion-qubit encoding (as JW is the most commonly used encoding). In orange, the TOPP-HATT optimised results are shown, where the same JW encoded Hamiltonian has been optimised with our method. The average reduction in circuit depth achieved by using the optimised encoding is $19.7\%$, with full details found in Table \ref{tab:water_depths} of Appendix \ref{appendix:water_depths}.






\begin{figure}[h!]
    \begin{subfigure}{0.49\textwidth}
        \includegraphics[width=\linewidth]{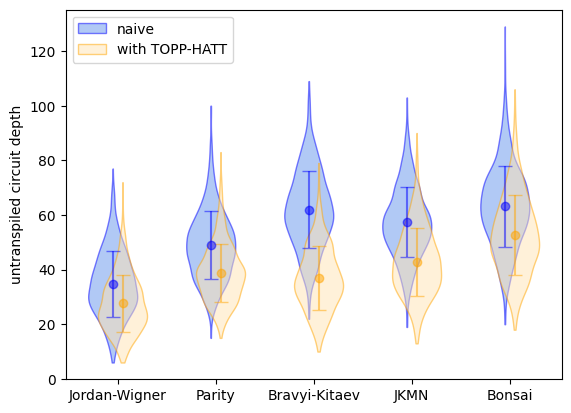}
        \caption{} \label{fig:untranspiled_qdrift}
    \end{subfigure}
    \hspace*{\fill}
    \begin{subfigure}{0.49\textwidth}
        \includegraphics[width=\linewidth]{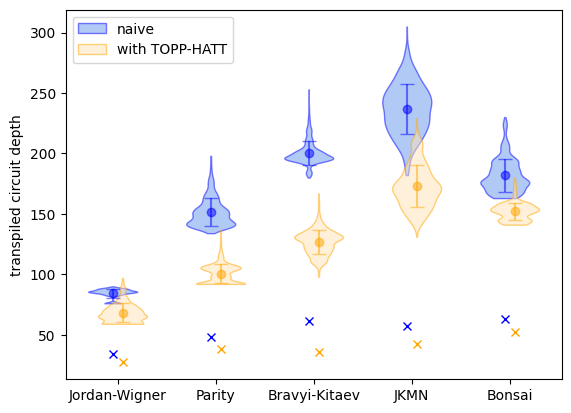}
        \caption{} \label{fig:transpiled_qdrift}
    \end{subfigure}
    \caption{qDRIFT circuit depths of \emph{STO-3G} water with fixed evolution duration 0.001 before (a), and after (b) transpilation for all the encodings studies in this work (with and without TOPP-HATT optimisation). In subfigure (a), 1000 qDRIFT circuits were constructed for each encoding. In subfigure (b), a circuit of mean depth from the initial construction was transpiled 1000 times using the \texttt{Qiskit} transpiler with optimisation level 3. In both cases, the circular markers indicate the mean value of these repeats, and the errorbars show one standard deviation. The cross markers in subfigure (b) show the depth of the input circuit to the transpiler. The circuits are transpiled using the 20-qubit IQM Garnet device topology. The Bonsai encoding is constructed using this topology and with the `heterogenous' labelling scheme.} \label{fig:qdrift_transpiling}
\end{figure}

Further, in Figure \ref{fig:qdrift_transpiling} we show the distribution of qDRIFT circuit depths for all encodings studied in this work, before and after transpilation to the 20-qubit IQM Garnet square-lattice device topology. Subfigure \ref{fig:untranspiled_qdrift} shows, for each encoding, the distribution of circuit depths for 1000 untranspiled circuits with an evolution duration of $0.001$. The average circuit depth is indicated by a circular marker and one standard deviation is shown by error bars. We see that for every encoding, TOPP-HATT optimisation reduces the mean circuit depth. The average reduction in circuit depth among all enocodings is $24.7\%$, with full details in Table \ref{tab:qdrift_mean_depths}.
We then take a circuit of mean depth for each of the encodings and pass this through the \texttt{Qiskit} circuit transpiler (with optimisation level 3) \cite{kremer_practical_2025}, the results of which are shown in Figure \ref{fig:transpiled_qdrift}. As the transpilation procedure is also stochastic, we pass each mean-depth qDRIFT circuit through the transpiler 1000 times, to produce a distribution of transpiled circuit depths. Again, the mean and standard deviation are shown by a circular marker and error bars. The depth of the input circuit is indicated with the cross markers. Transpilation naturally increases circuit depth due to the restrictions of device topology, but we find that again the TOPP-HATT optimisation yields shallower circuits across the set of encodings, on average we find a reduction of $26.5\%$, with full details in Table \ref{tab:qdrift_mean_depths}.

\begin{table}[t!]
    \centering
    \setlength{\tabcolsep}{5pt}
    \begin{tabular}{|c|ccc|ccc|}
    \hline \hline
                  & \multicolumn{3}{c|}{untranspiled depth}                                      & \multicolumn{3}{c|}{transpiled depth}                                        \\ 
    encoding      & \multicolumn{1}{c|}{naive} & \multicolumn{1}{c|}{optimised} & reduction (\%) & \multicolumn{1}{c|}{naive} & \multicolumn{1}{c|}{optimised} & reduction (\%) \\ 
    \hline
    Jordan-Wigner & \multicolumn{1}{c|}{34.8}  & \multicolumn{1}{c|}{27.7}      & 20.28           & \multicolumn{1}{c|}{84.3}  & \multicolumn{1}{c|}{68.3}      & 19.01           \\ 
    Parity        & \multicolumn{1}{c|}{48.9}  & \multicolumn{1}{c|}{38.8}      & 20.62           & \multicolumn{1}{c|}{151.2} & \multicolumn{1}{c|}{100.5}      & 33.5           \\ 
    Bravyi-Kitaev & \multicolumn{1}{c|}{62.0}  & \multicolumn{1}{c|}{36.9}      & 40.47           & \multicolumn{1}{c|}{200.5} & \multicolumn{1}{c|}{126.9}     & 36.72           \\ 
    JKMN          & \multicolumn{1}{c|}{57.4}  & \multicolumn{1}{c|}{42.9}      & 25.31           & \multicolumn{1}{c|}{236.5} & \multicolumn{1}{c|}{172.7}     & 26.98           \\ 
    Bonsai        & \multicolumn{1}{c|}{63.2}  & \multicolumn{1}{c|}{52.7}      & 16.68           & \multicolumn{1}{c|}{181.8} & \multicolumn{1}{c|}{152.2}     & 16.31           \\ 
    \hline
    mean          & \multicolumn{1}{c|}{}      & \multicolumn{1}{c|}{}          & \textbf{24.7}           & \multicolumn{1}{c|}{}      & \multicolumn{1}{c|}{}          & \textbf{26.5}           \\ 
    \hline \hline
    \end{tabular}
    \caption{Untranspiled and transpiled mean qDRIFT circuit depths for a batch of 1000 circuits for water in STO-3G basis at evolution duration 0.001. Depths for circuits using naive and TOPP-HATT optimised encodings are shown, along with the reduction in depth from using the optimised encoding as a percentage.}
    \label{tab:qdrift_mean_depths}
\end{table}

\subsection{Computation Time and Scaling}
\label{sec:runtime}
\begin{figure}[h!]
    \centering
    \begin{subfigure}[b]{0.45\textwidth}
        \includegraphics[width=\textwidth]{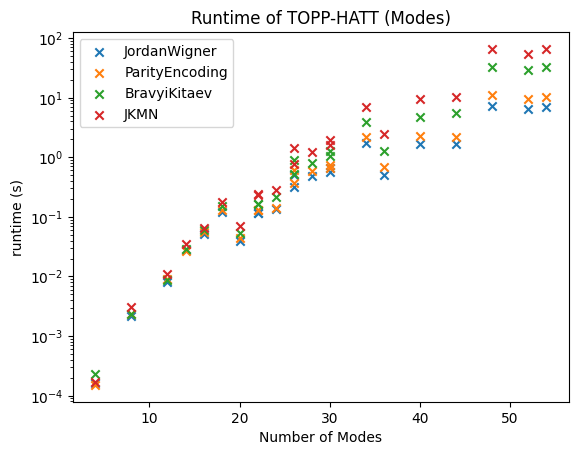}
        \caption{Number of modes}
    \end{subfigure}
    \begin{subfigure}[b]{0.45\textwidth}
        \includegraphics[width=\textwidth]{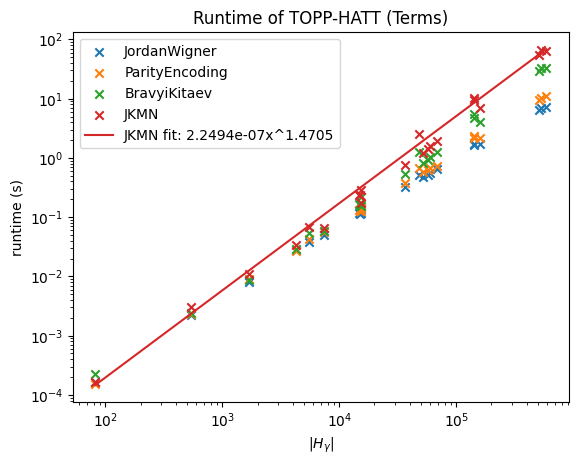}
        \caption{Number of Hamiltonian terms}
    \end{subfigure}
    \caption{Single-core computation time for encodings of small molecules in \emph{STO-3G} and \emph{6-31G} orbital basis sets, described by Table \ref{tab:smolmol}. a) The \emph{x}-axis shows the number of fermionic modes in the encoded Hamiltonian. b) The \emph{x}-axis shows the number of Hamiltonian terms $|H_{\gamma}|$. Each shows the Jordan-Wigner (green), Parity (red), Bravyi-Kitaev (blue) and JKMN (orange) encodings. Numerical results are shown in Table \ref{tab:smolmol}.}
    \label{fig:runtime}
\end{figure}

The computation time of our method is dependent on several factors; the number of fermionic modes in the target Hamiltonian, the number of terms in the Hamiltonian, and the structure of the tree encoding which is optimised. Figure \ref{fig:runtime} shows the single-core runtime for TOPP-HATT when applied to a variety of small molecules. Details and numerical results are given in Table \ref{tab:smolmol} of Appendix \ref{appendix:runtime}. Worst-case runtimes are found for the JKMN encoding as multiple nodes may be \emph{active} for each iteration, fitting these results shows runtime scales with $|H_{\gamma}|^{1.47}$. We note also that the inner loop of the algorithm (beginning line 13 of \ref{alg:TOPP-HATT}) is highly parallelizable.

We do not include in these figures the time required to prepare a fermionic Hamiltonian from electron-integrals, nor do we include the time required to encode a Hamiltonian using the optimised encoding determined by our method, given that both of these steps are required to make use of an unoptimised encoding.

\section{Conclusion}
Our method exhibits consistent reduction in Pauli-weight and coefficient-scaled Pauli-weight for a variety of standard encodings. Further, we have demonstrated improved performance when TOPP-HATT is applied to TTs derived from existing methods which optimise over the Hamiltonian. The combination of our method with trees derived from device connectivity shows reduction in both costs.

Applying our method as a pre-processing step for the Hamiltonian of water in the \emph{STO-3G} basis, before passing this to the qDRIFT algorithm at fixed evolution duration shows reductions in circuit depth by, on average, $24.7\%$ for the untranspiled case and $26.5\%$ for the transpiled case. We also see a reduction in circuit depth of $19.5\%$ over a larger range of evolution times.

Given the general applicability and low classical resource cost of this method, we anticipate that its use will benefit a broad range of gate-based quantum simulation methods.

\ack{
We thank LRZ for their support and facilitating access to high performance compute, in addition to the IQM \texttt{QExa20} superconducting device.
}

\funding{
MWdlB and TMB acknowledge support from the Engineering and Physical Sciences Research Council (EPSRC, grant numbers EP/S021582/1, EP/T517793/1 and EP/W524335/1). 
PVC is grateful for funding from the European Commission for VECMA (800925) and EPSRC for SEAVEA (EP/W007711/1).
We thank IQM for providing access to their superconducting devices.
}

\roles{MIWdlB: Conceptualization, Methodology, Software, Validation, Investigation, Writing - Original Draft, Visualisation.
TMB: Validation, Investigation, Writing - Original Draft, Visualisation.
PVC: Writing - Review \& Editing, Supervision, Project administration, Funding acquisition.}

\data{All source code, including scripts used to generate results presented in this paper, are freely available on our GitHub repository \href{https://github.com/UCL-CCS/ferrmion}{https://github.com/UCL-CCS/ferrmion}, and have been archived on Zenodo \href{https://zenodo.org/records/17407352}{https://zenodo.org/records/17407352}. Molecule geometries were obtained from \texttt{PubChem}, with electron integrals prepared by \texttt{PySCF} and \texttt{Openfermion}.}



\appendix
\section{qDRIFT Circuit Depth: Numerical Results}
\label{appendix:water_depths}

Table \ref{tab:water_depths} below gives the numerical results used to generate Figure \ref{fig:qdrift_depths_water}.

\begin{table}[h!]
    \centering
    \setlength{\tabcolsep}{10pt}
    \begin{tabular}{|c|c|c|c|}
    \hline \hline
    $t$      & naive  & optimised & reduction (\%) \\
    \hline
    0.0010 & 34.1 & 27.9 & 18.1  \\
    0.0013 & 53.3 & 42.3 & 20.6  \\
    0.0017 & 86.2 & 68.3 & 20.8  \\
    0.0022 & 140.5 & 112.2 & 20.2  \\
    0.0028 & 227.9 & 182.8 & 19.8  \\
    0.0036 & 380.1 & 305.7 & 19.6  \\
    0.0046 & 631.6 & 507.4 & 19.7  \\
    0.0060 & 1042.2 & 842.2 & 19.2  \\
    0.0077 & 1751.4 & 1409.6 & 19.5  \\
    0.0100 & 2930.8 & 2352.6 & 19.7  \\
    \hline
    mean   &        &           & \textbf{19.7}           \\
    \hline \hline
    \end{tabular}
    \caption{Mean untranspiled qDRIFT circuit depths with naive Jordan-Wigner and TOPP-HATT optimised Jordan-Wigner for water in \emph{STO-3G} basis over increasing evolution durations $t$, along with the reduction in depth from using the optimised encoding as a percentage.}
    \label{tab:water_depths}
\end{table}

\section{Runtime: Numerical Results}
\label{appendix:runtime}

Table \ref{tab:smolmol} below gives further details and numerical results for the results presented in Figure \ref{fig:runtime} of Section \ref{sec:runtime}.

\begin{table}[h!]
    \centering
    \small
    \setlength{\tabcolsep}{5pt}
    \begin{tabular}{|c|c|c|c|c|c|c|c|}
    \hline\hline
        Molecule& Basis & M& $|H_{\gamma}|$ & Jordan-Wigner & Parity & Bravyi-Kitaev & JKMN \\
    \hline
    \ce{H2}& STO-3G & 4 & 82 & 1.65e-4 (2.53e-5) & 1.51e-4 (4.30e-6) & 2.28e-4 (8.25e-5) & 1.68e-4 (1.56e-5) \\
    \ce{H2}& 6-31G & 8 & 544 & 2.21e-3 (7.15e-5) & 2.35e-3 (1.36e-4) & 2.37e-3 (1.48e-4) & 3.03e-3 (1.05e-3) \\
    \ce{LIH}& STO-3G & 12 & 1698 & 8.11e-3 (2.61e-4) & 9.02e-3 (4.32e-4) & 8.85e-3 (1.63e-4) & 1.08e-2 (5.38e-4) \\
    \ce{H2O}& STO-3G & 14 & 4263 & 2.78e-2 (1.65e-3) & 2.72e-2 (9.81e-4) & 2.94e-2 (4.82e-4) & 3.45e-2 (1.00e-3) \\
    \ce{BH3}& STO-3G & 16 & 7348 & 5.06e-2 (5.82e-3) & 5.55e-2 (4.23e-3) & 5.95e-2 (1.39e-3) & 6.62e-2 (7.52e-4) \\
    \ce{CH4}& STO-3G & 18 & 15293 & 1.18e-1 (1.67e-3) & 1.29e-1 (9.26e-4) & 1.53e-1 (5.65e-3) & 1.75e-1 (2.03e-3) \\
    \ce{N2}& STO-3G & 20 & 5510 & 3.87e-2 (5.05e-4) & 4.34e-2 (6.46e-4) & 5.33e-2 (1.78e-3) & 6.93e-2 (1.36e-3) \\
    \ce{HCN}& STO-3G & 22 & 14827 & 1.16e-1 (1.49e-3) & 1.32e-1 (2.68e-3) & 1.65e-1 (8.07e-3) & 2.33e-1 (1.25e-3) \\
    \ce{LIH}& 6-31G & 22 & 14923 & 1.15e-1 (2.62e-3) & 1.28e-1 (1.86e-3) & 1.62e-1 (2.67e-3) & 2.41e-1 (1.57e-2) \\
    \ce{C2H2}& STO-3G & 24 & 15360 & 1.35e-1 (1.70e-2) & 1.41e-1 (1.87e-3) & 2.17e-1 (4.50e-2) & 2.86e-1 (2.61e-3) \\
    \ce{H2O}& 6-31G & 26 & 36641 & 3.21e-1 (1.50e-2) & 3.72e-1 (1.46e-2) & 5.28e-1 (1.27e-2) & 7.64e-1 (1.08e-2) \\
    \ce{CH3F}& STO-3G & 26 & 57381 & 5.13e-1 (3.77e-3) & 6.17e-1 (7.85e-3) & 9.09e-1 (5.68e-3) & 1.44 (8.57e-3) \\
    \ce{ethene}& STO-3G & 28 & 51762 & 4.81e-1 (1.69e-2) & 5.79e-1 (6.82e-2) & 8.11e-1 (5.65e-3) & 1.22 (6.87e-3) \\
    \ce{ozone}& STO-3G & 30 & 60223 & 5.62e-1 (5.82e-3) & 6.70e-1 (4.09e-2) & 1.06 (2.24e-2) & 1.61 (3.45e-2) \\
    \ce{BH3}& 6-31G & 30 & 69163 & 6.51e-1 (1.38e-2) & 7.31e-1 (6.03e-3) & 1.26 (8.51e-2) & 1.93 (2.68e-2) \\
    \ce{CH4}& 6-31G & 34 & 160941 & 1.71 (3.05e-2) & 2.20 (4.56e-2) & 3.96 (4.04e-2) & 6.87 (6.33e-2) \\
    \ce{N2}& 6-31G & 36 & 48022 & 5.07e-1 (3.30e-2) & 6.86e-1 (7.90e-2) & 1.27 (2.09e-2) & 2.49 (6.05e-2) \\
    \ce{HCN}& 6-31G & 40 & 141192 & 1.69 (1.11e-1) & 2.31 (1.36e-1) & 4.66 (1.06e-1) & 9.47 (1.10e-1) \\
    \ce{C2H2}& 6-31G & 44 & 142218 & 1.66 (4.39e-2) & 2.20 (3.27e-2) & 5.59 (6.54e-2) & 1.01e1 (5.64e-2) \\
    \ce{CH3F}& 6-31G & 48 & 589652 & 7.25 (1.19e-1) & 1.13e1 (2.97e-1) & 3.28e1 (5.92e-1) & 6.47e1 (5.64e-1) \\
    \ce{ethene}& 6-31G & 52 & 514762 & 6.53 (1.61e-1) & 9.32 (3.75e-1) & 2.88e1 (4.58e-1) & 5.49e1 (6.20e-1) \\
    \ce{ozone}& 6-31G & 54 & 531319 & 6.91 (1.63e-1) & 1.01e1 (4.05e-2) & 3.25e1 (1.38e-1) & 6.57e1 (4.17e-1) \\
    \hline\hline
    \end{tabular}
    \caption{Runtime data for small molecules, as shown in \ref{fig:runtime}. $M$ gives the number of fermionic modes, $|H_{\gamma}|$ gives the number of Hamiltonian terms.  For each encoding, ten runs were performed with values given as the mean mean followed by (standard deviation).}
    \label{tab:smolmol}
\end{table}

\section{Other Molecules and Bases}
\begin{figure}[h!]
    \centering
    \includegraphics[width=\linewidth]{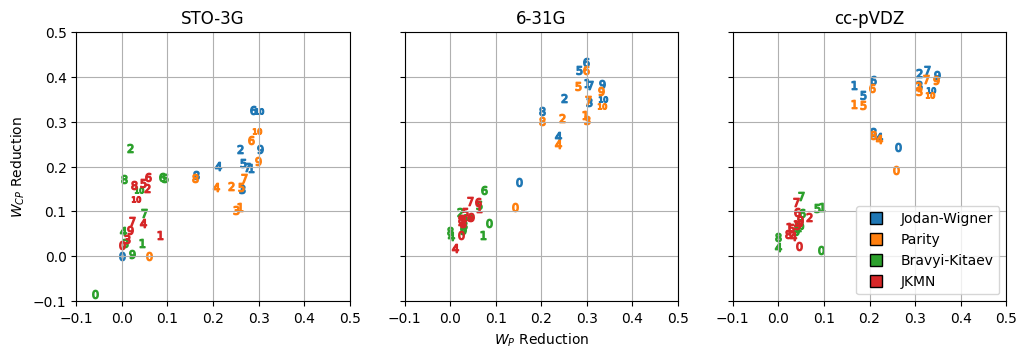}
    \caption{Reduction in $W_P$ and $W_{CP}$ resulting from TOPP-HATT, as a proportion of the naive encoding weight.}
    \label{fig:reduction}
\end{figure}

With the main results relating to water in $STO-3G$ basis, we show here that this method provides benefit when applied to arbitrary molecules and basis sets. Figure \ref{fig:reduction} shows the relative reduction in $W_P$ and $W_{CP}$ obtained by applying TOPP-HATT to molecules in the $STO-3G$, $6-31G$ and $cc-pVDZ$ bases. Each number used as a marker corresponds to a molecule, the key for this is given in Table \ref{tab:reductionkey}. All values are given as a percentage reduction of the weight obtained by applying the naive encoding. Each subfigure shows results for one basis. Note that the greatest benefit is obtained for the linear Jordan-Wigner and Parity encodings. These are also the least costly encodings to optimise.

\begin{table}[h!]
    \centering
    \begin{tabular}{|c|c|}
        \hline
        \hline
        Number & Molecule \\
        \hline
        0 & H2\\
        1 & LiH\\
        2 & H2O\\
        3 & BH3\\
        4 & CH4\\
        5 & N2\\
        6 & HCN\\
        7 & C2H2\\
        8 & CH3F\\
        9 & ethene\\
        10 & ozone\\
        \hline
        \hline
    \end{tabular}
    \caption{Key for Figure \ref{fig:reduction}}
    \label{tab:reductionkey}
\end{table}




\bibliography{references}

@misc{BuzzStrategicPaths2024,
	title = {Beyond the {Buzz}: {Strategic} {Paths} for {Enabling} {Useful} {NISQ} {Applications}},
	shorttitle = {Beyond the {Buzz}},
	url = {http://arxiv.org/abs/2405.14561},
	urldate = {2024-06-06},
	publisher = {arXiv},
	author = {Hegde, Pratibha Raghupati and Kyriienko, Oleksandr and Heimonen, Hermanni and Tolias, Panagiotis and Netzer, Gilbert and Barkoutsos, Panagiotis and Vinuesa, Ricardo and Peng, Ivy and Markidis, Stefano},
	month = may,
	year = {2024},
	note = {arXiv:2405.14561 [quant-ph]
Read\_Status: New
Read\_Status\_Date: 2024-11-11T13:35:59.504Z},
}

@article{JastrowtypeDecompositionQuantum2020,
	title = {Jastrow-type {Decomposition} in {Quantum} {Chemistry} for {Low}-{Depth} {Quantum} {Circuits}},
	volume = {16},
	issn = {1549-9618},
	url = {https://doi.org/10.1021/acs.jctc.9b00963},
	doi = {10.1021/acs.jctc.9b00963},
	number = {2},
	urldate = {2024-11-05},
	journal = {Journal of Chemical Theory and Computation},
	author = {Matsuzawa, Yuta and Kurashige, Yuki},
	month = feb,
	year = {2020},
	note = {Publisher: American Chemical Society
Read\_Status: New
Read\_Status\_Date: 2024-11-11T13:35:53.891Z},
	pages = {944--952},
}

@misc{bravyi_tapering_2017,
	title = {Tapering off qubits to simulate fermionic {Hamiltonians}},
	url = {http://arxiv.org/abs/1701.08213},
	urldate = {2025-12-15},
	publisher = {arXiv},
	author = {Bravyi, Sergey and Gambetta, Jay M. and Mezzacapo, Antonio and Temme, Kristan},
	month = jan,
	year = {2017},
	doi = {10.48550/arXiv.1701.08213},
}

@misc{mingare_tn4qa_2025,
	title = {{TN4QA}},
	url = {https://github.com/UCL-CCS/TN4QA},
	author = {Mingare, Angus and Heuzé, Isabelle},
	year = {2025},
	note = {https://github.com/UCL-CCS/TN4QA},
}

@article{li_huffman-code-based_2025,
	title = {Huffman-{Code}-{Based} {Ternary} {Tree} {Transformation}},
	volume = {42},
	issn = {0256-307X},
	url = {https://doi.org/10.1088/0256-307X/42/10/100001},
	doi = {10.1088/0256-307X/42/10/100001},
	number = {10},
	urldate = {2025-11-28},
	journal = {Chinese Physics Letters},
	author = {Li, Qing-Song and Liu, Huan-Yu and Wang, Qingchun and Wu, Yu-Chun and Guo, Guo-Ping},
	month = sep,
	year = {2025},
	pages = {100001},
}

@misc{kremer_practical_2025,
	title = {Practical and efficient quantum circuit synthesis and transpiling with {Reinforcement} {Learning}},
	url = {http://arxiv.org/abs/2405.13196},
	urldate = {2025-11-26},
	publisher = {arXiv},
	author = {Kremer, David and Villar, Victor and Paik, Hanhee and Duran, Ivan and Faro, Ismael and Cruz-Benito, Juan},
	month = feb,
	year = {2025},
	doi = {10.48550/arXiv.2405.13196},
}

@article{campbell_random_2019,
	title = {Random {Compiler} for {Fast} {Hamiltonian} {Simulation}},
	volume = {123},
	url = {https://link.aps.org/doi/10.1103/PhysRevLett.123.070503},
	doi = {10.1103/PhysRevLett.123.070503},
	number = {7},
	urldate = {2025-11-26},
	journal = {Physical Review Letters},
	author = {Campbell, Earl},
	month = aug,
	year = {2019},
	pages = {070503},
}

@article{ralli_implementation_2021,
	title = {Implementation of measurement reduction for the variational quantum eigensolver},
	volume = {3},
	url = {https://link.aps.org/doi/10.1103/PhysRevResearch.3.033195},
	doi = {10.1103/PhysRevResearch.3.033195},
	number = {3},
	urldate = {2025-11-18},
	journal = {Physical Review Research},
	author = {Ralli, Alexis and Love, Peter J. and Tranter, Andrew and Coveney, Peter V.},
	month = aug,
	year = {2021},
	pages = {033195},
}

@article{ReducingEntanglementPhysically2024,
  title = {Reducing {{Entanglement}} with {{Physically Inspired Fermion-To-Qubit Mappings}}},
  author = {{Parella-Dilm{\'e}}, Teodor and Kottmann, Korbinian and Zambrano, Leonardo and Mortimer, Luke and Kottmann, Jakob S. and Ac{\'i}n, Antonio},
  year = 2024,
  month = aug,
  journal = {PRX Quantum},
  volume = {5},
  number = {3},
  pages = {030333},
  issn = {2691-3399},
  doi = {10.1103/PRXQuantum.5.030333},
  urldate = {2024-11-07}
}

@inproceedings{liu_hatt_2025,
	title = {{HATT}: {Hamiltonian} {Adaptive} {Ternary} {Tree} for {Optimizing} {Fermion}-to-{Qubit} {Mapping}},
	shorttitle = {{HATT}},
	url = {http://arxiv.org/abs/2409.02010},
	doi = {10.1109/HPCA61900.2025.00022},
	urldate = {2025-04-28},
	booktitle = {2025 {IEEE} {International} {Symposium} on {High} {Performance} {Computer} {Architecture} ({HPCA})},
	author = {Liu, Yuhao and Yao, Kevin and Hong, Jonathan and Froustey, Julien and Rrapaj, Ermal and Iancu, Costin and Li, Gushu and Shi, Yunong},
	month = mar,
	year = {2025},
	pages = {143--157},
}

@misc{leimkuhler_exponential_2025,
	title = {Exponential quantum speedups in quantum chemistry with linear depth},
	url = {http://arxiv.org/abs/2503.21041},
	urldate = {2025-03-30},
	publisher = {arXiv},
	author = {Leimkuhler, Oskar and Whaley, K. Birgitta},
	month = mar,
	year = {2025},
	doi = {10.48550/arXiv.2503.21041},
}

@article{eriksen_shape_2021,
	title = {The {Shape} of {Full} {Configuration} {Interaction} to {Come}},
	volume = {12},
	issn = {1948-7185, 1948-7185},
	url = {http://arxiv.org/abs/2010.12304},
	doi = {10.1021/acs.jpclett.0c03225},
	number = {1},
	urldate = {2025-03-27},
	journal = {The Journal of Physical Chemistry Letters},
	author = {Eriksen, Janus J.},
	month = jan,
	year = {2021},
	pages = {418--432},
}

@article{jiang_optimal_2020,
	title = {Optimal fermion-to-qubit mapping via ternary trees with applications to reduced quantum states learning},
	volume = {4},
	doi = {10.22331/q-2020-06-04-276},
	urldate = {2024-05-01},
	journal = {Quantum},
	author = {Jiang, Zhang and Kalev, Amir and Mruczkiewicz, Wojciech and Neven, Hartmut},
	month = jun,
	year = {2020},
	pages = {276},
}

@misc{yu_clifford_2025,
	title = {Clifford circuit based heuristic optimization of fermion-to-qubit mappings},
	url = {http://arxiv.org/abs/2502.11933},
	urldate = {2025-03-26},
	publisher = {arXiv},
	author = {Yu, Jeffery and Liu, Yuan and Sugiura, Sho and Voorhis, Troy Van and Zeytinoğlu, Sina},
	month = feb,
	year = {2025},
	doi = {10.48550/arXiv.2502.11933},
}

@article{jordan_uber_1928,
	title = {Über das {Paulische} äquivalenzverbot},
	volume = {47},
	issn = {0044-3328},
	url = {https://doi.org/10.1007/BF01331938},
	doi = {10.1007/BF01331938},
	number = {9},
	urldate = {2025-02-25},
	journal = {Zeitschrift für Physik},
	author = {Jordan, P. and Wigner, E.},
	month = sep,
	year = {1928},
	pages = {631--651},
}

@misc{chiew_ternary_2024,
	title = {Ternary tree transformations are equivalent to linear encodings of the {Fock} basis},
	url = {http://arxiv.org/abs/2412.07578},
	urldate = {2024-12-17},
	publisher = {arXiv},
	author = {Chiew, Mitchell and Harrison, Brent and Strelchuk, Sergii},
	month = dec,
	year = {2024},
	doi = {10.48550/arXiv.2412.07578},
}

@misc{chien_optimizing_2022,
	title = {Optimizing fermionic encodings for both {Hamiltonian} and hardware},
	url = {http://arxiv.org/abs/2210.05652},
	urldate = {2024-03-26},
	publisher = {arXiv},
	author = {Chien, Riley W. and Klassen, Joel},
	month = oct,
	year = {2022},
	doi = {10.48550/arXiv.2210.05652},
}

@article{bravyi_fermionic_2002,
	title = {Fermionic quantum computation},
	volume = {298},
	issn = {00034916},
	url = {http://arxiv.org/abs/quant-ph/0003137},
	doi = {10.1006/aphy.2002.6254},
	number = {1},
	urldate = {2024-03-28},
	journal = {Annals of Physics},
	author = {Bravyi, Sergey and Kitaev, Alexei},
	month = may,
	year = {2002},
	pages = {210--226},
}

@misc{miller_treespilation_2024,
	title = {Treespilation: {Architecture}- and {State}-{Optimised} {Fermion}-to-{Qubit} {Mappings}},
	shorttitle = {Treespilation},
	url = {http://arxiv.org/abs/2403.03992},
	urldate = {2024-05-13},
	publisher = {arXiv},
	author = {Miller, Aaron and Glos, Adam and Zimborás, Zoltán},
	month = mar,
	year = {2024},
	doi = {10.48550/arXiv.2403.03992},
}

@article{miller_bonsai_2023,
	title = {Bonsai {Algorithm}: {Grow} {Your} {Own} {Fermion}-to-{Qubit} {Mappings}},
	volume = {4},
	shorttitle = {Bonsai {Algorithm}},
	url = {https://link.aps.org/doi/10.1103/PRXQuantum.4.030314},
	doi = {10.1103/PRXQuantum.4.030314},
	number = {3},
	urldate = {2024-03-26},
	journal = {PRX Quantum},
	author = {Miller, Aaron and Zimborás, Zoltán and Knecht, Stefan and Maniscalco, Sabrina and García-Pérez, Guillermo},
	month = aug,
	year = {2023},
	pages = {030314},
}

@book{nielsen_quantum_2012,
	series = {Quantum {Computation} and {Quantum} {Information}},
	title = {Quantum {Computation} and {Quantum} {Information}},
	isbn = {978-1-107-00217-3},
	url = {https://www.cambridge.org/core/product/identifier/9780511976667/type/book},
	publisher = {Cambridge University Press},
	author = {Nielsen, Michael A. and Chuang, Isaac L.},
	month = jun,
	year = {2012},
	doi = {10.1017/CBO9780511976667},
}

@article{feynman_simulating_1982,
	title = {Simulating physics with computers},
	doi = {10.1007/BF02650179},
	journal = {International Journal of Theoretical Physics},
	author = {Feynman, Richard P.},
	year = {1982},
}
\bibliographystyle{abbrv}

%
%

\end{document}